\documentclass[11pt,twoside,a4paper,cmspaper,final,collab]{cms-tdr}
\def\svnVersion{3ccf85d}\def\svnDate{2026/03/19}\def\cmsCernNoTag{CERN-EP-2025-143}\def\cmsCernDate{\today}\def\cmsMessage{Published in the Journal of High Energy Physics as \href{http://dx.doi.org/10.1007/JHEP03(2026)140}{\doi{10.1007/JHEP03(2026)140}.}}
\begin{document}\cmsNoteHeader{TOP-23-009}

\newlength\cmsFigWidth
\ifthenelse{\boolean{cms@external}}{\setlength\cmsFigWidth{0.45\textwidth}}{\setlength\cmsFigWidth{0.49\textwidth}}
\providecommand{\cmsLeft}{left\xspace}
\providecommand{\cmsRight}{right\xspace}

\newcommand{\antikt}{anti-\kt}
\newcommand{\pp}{\ensuremath{\Pp\Pp}\xspace}
\newcommand{\fb}{\unit{fb}}
\newcommand{\mb}{\unit{mb}}
\newcommand{\Wjets}{\ensuremath{\PW\text{+jets}}\xspace}
\newcommand{\WZ}{\ensuremath{\PW\PZ}\xspace}
\newcommand{\ZZ}{\ensuremath{\PZ\PZ}\xspace}
\newcommand{\ttZ}{\ensuremath{\PQt\PAQt\PZ}\xspace}
\newcommand{\ttgamma}{\ensuremath{\PQt\PAQt\PGg}\xspace}
\newcommand{\ttX}{\ensuremath{\PQt\PAQt\PX}\xspace}
\newcommand{\tWZ}{\ensuremath{\PQt\PW\PZ}\xspace}
\newcommand{\ttW}{\ensuremath{\PQt\PAQt\PW}\xspace}
\newcommand{\ttH}{\ensuremath{\PQt\PAQt\PH}\xspace}
\newcommand{\tZq}{\ensuremath{\PQt\PZ\PQq}\xspace}
\newcommand{\tttt}{\ensuremath{\PQt\PAQt\PQt\PAQt}\xspace}
\newcommand{\SRWZ}{\ensuremath{\text{SR}_{\WZ}}\xspace}
\newcommand{\SRZZ}{\ensuremath{\text{SR}_{\ZZ}}\xspace}
\newcommand{\SRttZ}{\ensuremath{\text{SR}_{\ttZ}}\xspace}
\newcommand{\CRWZ}{\ensuremath{\text{CR}_{\WZ}}\xspace}
\newcommand{\CRttZ}{\ensuremath{\text{CR}_{\ttZ}}\xspace}
\newcommand{\VRWZ}{\ensuremath{\text{VR}_{\WZ}}\xspace}
\newcommand{\VRttZ}{\ensuremath{\text{VR}_{\ttZ}}\xspace}
\newcommand{\mll}{\ensuremath{m_{\Pell\Pell}}\xspace}
\newcommand{\Singlet}{\ensuremath{\mathcal{O}_{\varphi q}^{(1)}}\xspace}
\newcommand{\Triplet}{\ensuremath{\mathcal{O}_{\varphi q}^{(3)}}\xspace}
\newcommand{\UpO}{\ensuremath{\mathcal{O}_{\varphi u}}\xspace}
\newcommand{\DownO}{\ensuremath{\mathcal{O}_{\varphi d}}\xspace}
\newcommand{\Singletab}{\ensuremath{\mathcal{O}_{\varphi q}^{(1)(ab)}}\xspace}
\newcommand{\Tripletab}{\ensuremath{\mathcal{O}_{\varphi q}^{(3)(ab)}}\xspace}
\newcommand{\UpOab}{\ensuremath{\mathcal{O}_{\varphi u}^{(ab)}}\xspace}
\newcommand{\DownOab}{\ensuremath{\mathcal{O}_{\varphi d}^{(ab)}}\xspace}
\newcommand{\cHqSinglet}{\ensuremath{c_{\varphi q}^{(1)}}\xspace}
\newcommand{\cHqTriplet}{\ensuremath{c_{\varphi q}^{(3)}}\xspace}
\newcommand{\cHqMinus}{\ensuremath{c_{\varphi q}^{(-)}}\xspace}
\newcommand{\cHqTripletLight}{\ensuremath{c_{\varphi q}^{(3)(11+22)}}\xspace}
\newcommand{\cHqTripletThird}{\ensuremath{c_{\varphi q}^{(3)(33)}}\xspace}
\newcommand{\cHqMinusLight}{\ensuremath{c_{\varphi q}^{(-)(11+22)}}\xspace}
\newcommand{\cHqMinusThird}{\ensuremath{c_{\varphi q}^{(-)(33)}}\xspace}
\newcommand{\cHu}{\ensuremath{c_{\varphi u}}\xspace}
\newcommand{\cHd}{\ensuremath{c_{\varphi d}}\xspace}
\newcommand{\cHuLight}{\ensuremath{c_{\varphi u}^{(11+22)}}\xspace}
\newcommand{\cHuThird}{\ensuremath{c_{\varphi u}^{(33)}}\xspace}
\newcommand{\cHdLight}{\ensuremath{c_{\varphi d}^{(11+22)}}\xspace}
\newcommand{\cHdThird}{\ensuremath{c_{\varphi d}^{(33)}}\xspace}
\newcommand{\cW}{\ensuremath{c_{W}}\xspace}
\newcommand{\cWtil}{\ensuremath{c_{\widetilde{W}}}\xspace}
\newcommand{\derivTwoSide}{\ensuremath{\overset{\leftrightarrow}{D}}\xspace}
\newcommand{\Nlep}{\ensuremath{N_\Pell}\xspace}

\newlength\cmsTabSkip\setlength{\cmsTabSkip}{1ex}
\cmsNoteHeader{FlavorEFT}
\title{Probing the flavour structure of dimension-6 EFT operators in multilepton final states in proton-proton collisions at \texorpdfstring{$\sqrt{s}=13\TeV$}{sqrt(s)=13 TeV} }

\author[cern]{The CMS Collaboration}

\date{\today}

\abstract{
An analysis of the flavour structure of dimension-6 effective field theory~(EFT) operators in multilepton final states is presented, focusing on the interactions of quarks with \PZ bosons. For the first time, the flavour structure of these operators is disentangled by simultaneously probing the interactions with different quark generations. The analysis targets the associated production of a top quark pair and a \PZ boson, as well as diboson processes in final states with at least three leptons, which can be electrons or muons. The data were recorded by the CMS experiment in the years 2016--2018 in proton-proton collisions at a centre-of-mass energy of 13\TeV and correspond to an integrated luminosity of 138\fbinv. Consistency with the standard model of particle physics is observed and limits are set on the selected Wilson coefficients, split into couplings to light- and heavy-quark generations.
}

\hypersetup{
pdfauthor={CMS Collaboration},
pdftitle={Probing the flavour structure of dimension-6 EFT operators in multilepton final states in proton-proton collisions at sqrt(s) = 13 TeV},
pdfsubject={CMS},
pdfkeywords={CMS, top quark, EFT, flavour}}

\maketitle

\section{Introduction}
\label{sec:introduction}
Despite the known incompleteness of the standard model (SM) of particle physics,
no evidence for new physics phenomena has been found in direct searches at the CERN LHC.
The reason may lie in the limited energy range of the LHC, which prevents beyond-the-SM~(BSM)
physics situated at much higher energy scales to be directly produced.
However, BSM physics could still leave traces in measurements at the LHC
in the form of deviations from the SM prediction.
In this case, an effective field theory~(EFT)~\cite{Burgess:2020tbq,Buchmuller:1985jz,Brivio:2017vri}
is a suitable framework for describing the effects of high-energy BSM phenomena at the energies that can be probed with the LHC.
An EFT is valid if the scale of BSM physics, $\Lambda$, is well above the energy scale probed by an experiment.
The standard model EFT~(SMEFT)~\cite{Buchmuller:1985jz,Giudice:2007fh,Grzadkowski:2010es} framework
extends the SM Lagrangian with newly introduced operators,
organised by their mass dimension.
With this, all possible BSM effects are incorporated in an effective Lagrangian,
which can be written as
\begin{equation}
\mathcal{L}_{\text{eff}} = \mathcal{L}_{\text{SM}} + \sum_{d,i} \frac{c_{i}}{\Lambda^{d-4}} \mathcal{O}_{i},
\end{equation}
where $\mathcal{L}_{\text{SM}}$ is the Lagrangian of the SM,
$d$ is the mass dimension,
and $\mathcal{O}_{i}$ are the newly introduced operators.
For each $\mathcal{O}_{i}$, there exists a dimensionless Wilson coefficient~(WC), $c_{i}$,
scaling the strength of the respective operator.
With the SM defined at $d=4$ and the suppression of the BSM physics with
$1/\Lambda^{d-4}$, low mass dimensions,
starting at $d=5$, should become observable first.
However, extensions to odd number mass dimensions are assumed to be suppressed as
sizeable contributions from these operators lead to the
violation of the conservation of lepton or baryon numbers~\cite{HELSET2020135132}.
Thus, in this analysis, only operators of mass dimension $d=6$ are considered.

A subset of these operators modify the couplings of electroweak~(EW) bosons to quarks
and can be probed in multiple processes at the LHC.
In the top quark sector, these couplings were studied in the context of EFT by the
ATLAS~\cite{ATLAS:2023eld,ATLAS:2024hmk} and CMS~\cite{CMS:2021aly,CMS:2022hjj,CMS:2020lrr,CMS:2023xyc}
Collaborations.
Even though many analyses in the CMS programme already probe EFT effects in
multiple processes simultaneously~\cite{CMS:2021aly,CMS:2022hjj,CMS:2020lrr,CMS:2023xyc}, they only consider the
couplings involving top and bottom quarks.
However, even in the case of the associated production of a top quark pair~(\ttbar) and a \PZ boson~(\ttZ), the \PZ boson
can be radiated from initial-state quarks such that the EW coupling
to light-quark generations becomes relevant.
In addition, couplings to the light-quark generations enter analyses focusing on top quark processes
through background processes such as \WZ production in \ttZ event selections.
In the analysis reported in this article, we probe the flavour structure of dimension-6 EFT couplings by
simultaneously measuring the quark couplings with a \PZ boson for different quark generations.
This extends the EFT programme at the LHC by incorporating a greater awareness of flavour dependencies, which are often not considered.
We aim to study the interactions between \PZ bosons and quarks, which are
also relevant in neutral-current transitions of \PQb to \PQs quarks
of the form $\PQb \to \PQs \Pellp\Pellm$~\cite{Bissmann:2020mfi,Bruggisser:2021duo},
where $\Pellp\Pellm$ denotes a charged-lepton pair.
The probed EFT operators are defined in the Warsaw basis~\cite{Grzadkowski:2010es} and the measurement
focuses on those that modify the EW vector couplings,
\begin{equation}
\begin{aligned}
  \Singletab &= \left(\varphi^{\dagger} i \derivTwoSide_{\mu} \varphi\right) \left(\overline{q_{a}} \gamma^\mu q_{b}\right), \\
  \Tripletab &= \left(\varphi^{\dagger} i \derivTwoSide_{\mu}^{I} \varphi\right) \left(\overline{q_{a}} \gamma^\mu \tau^{I} q_{b}\right), \\
  \UpOab     &= \left(\varphi^{\dagger} i \derivTwoSide_{\mu} \varphi\right) \left(\overline{u_{a}} \gamma^\mu u_{b} \right),\ \text{and} \\
  \DownOab   &= \left(\varphi^{\dagger} i \derivTwoSide_{\mu} \varphi\right) \left(\overline{d_{a}} \gamma^\mu d_{b} \right),
\end{aligned}
\end{equation}
where $\varphi$ is the Higgs field, $q$ is the left-handed quark field,
$u$ and $d$ are the right-handed quark fields, the indices $a$ and $b$ define the quark flavour,
\derivTwoSide is a two-sided covariant derivative,
$\gamma^\mu$ are the gamma matrices, and $\tau^I$ are the Pauli sigma matrices.
Here, we focus on the diagonal entries of the operators (where $a=b$), such that only processes are considered
in which the quark generation remains unchanged after the interaction with an
EW boson.
Off-diagonal entries would introduce flavour-changing neutral currents, which are tightly constrained
and are found to have negligible effect on the selected processes~\cite{ATLAS:2018zsq,ATLAS:2021amo,ATLAS:2022per,CMS:2019zct}.
The coupling of the \PZ boson to quarks is modified by combinations of the respective
WCs \cHqSinglet and \cHqTriplet,
while the \PW boson interaction changes with \cHqTriplet.
The right-handed operators \UpO and \DownO are similarly split into the three quark flavours
and only modify the coupling to the \PZ boson.
Similar to existing EFT interpretations~\cite{CMS:2021aly, CMS:2023xyc}, the probed WCs are reparametrised in the basis of $\cHqMinus = \cHqSinglet-\cHqTriplet$
and a redefined \cHqTriplet such that, by adjusting \cHqSinglet accordingly, \cHqMinus remains constant when altering \cHqTriplet.
This basis allows a more direct mapping of the operators to the physical couplings of the \PZ and \PW bosons.

\begin{figure*}[tb]
  \centering
    \includegraphics[width=0.98\textwidth]{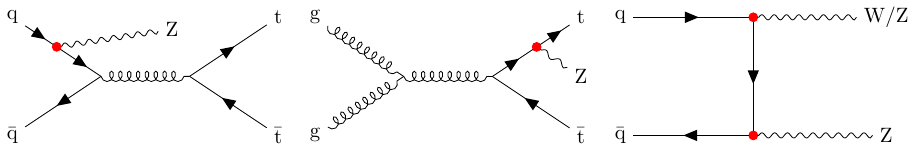}
    \caption{Representative Feynman diagrams showing the leading-order contributions to the \ttZ production,
    with the \PZ boson radiated from the initial-state quarks~(left) and from one of the top quarks~(middle).
    The \WZ and/or \ZZ production is also shown~(right).
    The vertices affected by the EFT operators probed in this analysis are highlighted
    with red dots.
    \label{fig:feynman}}
\end{figure*}

In order to achieve sensitivity to the flavour structure of the \Singlet and \Triplet
operators, multiple processes with similar multilepton final states are considered.
A simultaneous study of \ttZ, \WZ, and \ZZ production
allows us to disentangle the coupling strengths to the third quark generations compared to lighter ones,
as shown in Fig.~\ref{fig:feynman}.
The \ttZ process gives sensitivity to the third-generation couplings due to diagrams
where the \PZ boson is radiated from one of the top quarks.
In addition, sensitivity to first- and second-generation couplings arises from
diagrams where the \PZ boson is radiated from initial-state quarks.
The \ZZ and \WZ processes increase the constraining power for the first and
second quark generations since the bosons are dominantly radiated from light quarks.
Since a $\PW\PW\PZ$ vertex can appear in the \WZ production process,
the operators
\begin{equation}
\begin{aligned}
\mathcal{O}_W               &= \epsilon^{IJK} W_\mu^{I \nu} W_\nu^{J \rho} W_\rho^{K \mu}\ \text{and}\\
\mathcal{O}_{\widetilde{W}} &= \epsilon^{IJK} \widetilde{W}_\mu^{I \nu} W_\nu^{J \rho} W_\rho^{K \mu},
\end{aligned}
\end{equation}
where $W$ is the \PW boson field and $\epsilon$ is the Levi-Civita symbol, are considered in this analysis as well.
We always target the leptonic decay of the \PZ boson and select
final states containing three or four electrons and muons, referred to simply as leptons from hereon.
This choice ensures a high-quality reconstruction of the \PZ boson, which is our
handle to provide a highly pure sample of signal processes.
A profiled likelihood fit to the distribution in the transverse momentum, \pt, of the reconstructed \PZ boson candidate is used to probe the EFT effects.
The fit is performed simultaneously in three signal regions~(SRs) that target the \ttZ, \WZ, and \ZZ processes.
Control regions~(CRs) are defined to estimate the background contribution from processes with nonprompt leptons.

In this analysis, proton-proton~(\pp) collision data are used, collected at $\sqrt{s}=13\TeV$
in 2016--2018 and corresponding to an integrated luminosity of 138\fbinv.
Dimension-6 EFT operators and the corresponding WCs \cHqMinus, \cHqTriplet, \cHu, \cHd, \cW, and \cWtil are probed.
Aiming towards a more global interpretation, the flavour-dependent interaction strengths
of the aforementioned EFT operators are simultaneously targeted in
\ttZ, \WZ, and \ZZ processes, accounting for correlations of EFT effects in these processes by construction.
Moreover, this approach enables a proper and consistent treatment of the
correlations between systematic uncertainties,
which is crucial for ensuring the reliability of the interpretations of the results.
Other operators such as $\mathcal{O}_{tZ}$ and $\mathcal{O}_{tW}$ are not considered as they are already strongly 
constrained in other analyses of \ttbar production in association with a \PZ boson~\cite{ATLAS:2023eld,CMS:2021aly,CMS:2022hjj,CMS:2020lrr,CMS:2023xyc}. 
In addition, measurements including the associated production of a \ttbar pair with a photon (\ttgamma)~\cite{ATLAS:2024hmk,CMS:2022lmh} 
provide even more stringent limits.

The paper is organised as follows. The CMS detector is described in Section~\ref{sec:detector} and
the simulated event samples are detailed in Section~\ref{sec:dataMC}.
Sections~\ref{sec:reconstruction} and~\ref{sec:selection} discuss the reconstruction
of physics objects and the event selection in the various regions of the analysis, respectively.
The background estimation of processes with nonprompt leptons is detailed in Section~\ref{sec:nonprompt}.
In Section~\ref{sec:uncertainties}, we explain sources
of systematic uncertainties and their estimation.
The profiled maximum likelihood fit to data and results are discussed in Section~\ref{sec:results} and,
lastly, a summary is provided in Section~\ref{sec:summary}.
Tabulated results are provided in the HEPData record for this analysis~\cite{hepdata}.

\section{The CMS detector}
\label{sec:detector}

The central feature of the CMS detector is a superconducting solenoid of 6\unit{m} internal diameter,
providing a magnetic field of 3.8\unit{T}. A silicon pixel and strip tracker, a lead tungstate crystal
electromagnetic calorimeter (ECAL), and a brass and scintillator hadron calorimeter (HCAL),
each composed of a central barrel and two endcap sections, reside within the solenoid volume.
Forward calorimeters extend the pseudorapidity ($\eta$) coverage provided by the barrel and endcap detectors.
Muons are detected in gas-ionisation chambers embedded in the steel flux-return yoke outside the solenoid.
A more detailed description of the CMS detector, together with a definition of the coordinate system,
can be found in Ref.~\cite{Chatrchyan:2008zzk}.
Between the 2016 and 2017 data-taking runs, the CMS pixel detector was upgraded with additional layers
in the barrel and endcap regions of the CMS detector.
Details about the changes can be found in Ref.~\cite{CMSTrackerGroup:2020edz}.

Events of interest are selected using a two-tiered trigger system. The first
level, composed of custom hardware processors, uses information from the
calorimeters and muon detectors to select events at a rate of around
100\unit{kHz} within a fixed latency of about 4\mus~\cite{CMS:2020cmk}. The
second level, known as the high-level trigger, consists of a farm of
processors running a version of the full event reconstruction software optimised
for fast processing, and reduces the event rate to a few kHz before
data storage~\cite{Khachatryan:2016bia,CMS:2024aqx}.

\section{Simulated event samples}
\label{sec:dataMC}

{\tolerance=800
We simulate all processes that are compatible with our selected final state and
apply the same reconstruction and event selection as for data.
The standard model predictions of the \ttZ, \WZ, and \ZZ processes are obtained by 
simulating events at next-to-leading order~(NLO) in perturbative quantum chromodynamics~(QCD).
We use \MGvATNLO v2.6.5~\cite{Alwall:2014hca, Frixione:2002ik} for \ttZ production and 
\POWHEG~v2~\cite{Melia:2011tj,Nason:2013ydw} for \WZ and \ZZ.
For the prediction of EFT effects, alternative samples of the \ttZ, \WZ, and \ZZ processes are simulated using \MGvATNLO 
at leading order~(LO) in QCD and in EW theory with up to one additional parton simulated in the matrix element calculations.
A reweighting method~\cite{Mattelaer:2016gcx} is used to reproduce the effect of different WC values on event distributions.
The total matrix elements are calculated at several points in the EFT parameter space
using the SMEFTsim package~\cite{Brivio:2017btx, Brivio:2020onw} via
\begin{equation}
\abs{\mathcal{M}}^{2} = \abs{\mathcal{M}_\text{SM}}^{2} + \sum_{i} 2 c_i \text{Re}(\mathcal{M}_\text{SM}^{*}\mathcal{M}_\text{BSM}^{i}) + \sum_{i}\sum_{j} c_i c_j \text{Re}(\mathcal{M}_\text{BSM}^{*i}\mathcal{M}_\text{BSM}^{j}),
\end{equation}
where $\mathcal{M}_\text{SM}$ and $\mathcal{M}_\text{BSM}$ are the matrix element contributions
from SM and BSM physics, respectively.
The WCs, $c_i$, are assumed to be real and the sum runs over all considered EFT operators.
Weights corresponding to the ratios $\abs{\mathcal{M}}^{2}/\abs{\mathcal{M}_\text{SM}}^{2}$
for selected base points in the EFT parameter space are stored for each simulated event.
With $(N+1)(N+2)/2$ of these weights,
the polynomial for $N$ EFT operators is solved to extract weights,
containing the full dependence on WCs, on an event-by-event basis.
In the computation of $\mathcal{M}$, we only consider diagrams with a single EFT vertex such that terms
linear in $c$ encode the SM-BSM interference and quadratic terms describe the pure BSM contribution.
Setting all WCs to zero, the LO samples with an additional parton are verified to reproduce---within
renormalisation and factorisation scale uncertainties estimated following the description in Section~\ref{sec:uncertainties}---the SM
distributions modelled at NLO in perturbative QCD.
The predictions at NLO are used for the SM scenario of \ttZ, \WZ, and \ZZ and
samples with EFT weights are only used to predict the EFT effects.
\par}

Top quark pair production, \ttbar production in association with a Higgs boson (\ttH),
and single top $t$-channel and $\PQt\PW$ processes are simulated at NLO in QCD using
\POWHEG~\cite{Nason:2004rx, Frixione:2007vw, Alioli:2010xd, Frixione:2007nw, Hartanto:2015uka, Alioli:2009je,Re:2010bp}.
The \MGvATNLO generator is used for the simulation of \ttW, \tttt,
\tWZ, \tZq, and triboson production ($\PW\PW\PW$, $\PW\PW\PZ$, $\PW\PZ\PZ$, and $\PZ\PZ\PZ$) at NLO in QCD.
The Drell--Yan process, \Wjets, and \ttbar production in association
with two bosons ($\PQt\PAQt\PW\PW$, $\PQt\PAQt\PW\PZ$, $\PQt\PAQt\PZ\PZ$) are simulated
with \MGvATNLO at LO.
In the following, we group the \ttW, \ttH, $\PQt\PAQt\PW\PW$, $\PQt\PAQt\PW\PZ$, $\PQt\PAQt\PZ\PZ$, and \tttt processes as \ttX.

The $\PW\PW$ process and gluon-initiated \ZZ production~($\Pg\Pg\to\ZZ$) are simulated at LO with
\PYTHIA~8.240~\cite{Sjostrand:2014zea} and the \MCFM v7.0.1 generator~\cite{Campbell:1999ah,Campbell:2011bn,Campbell:2015qma}, respectively.
Other processes, such as the production of \ttgamma,
the simultaneous production of a \PZ boson and a photon,
additional processes with a Higgs boson decaying to a pair of \PZ bosons, and the
purely EW production of $\PQt\PAQt\PW$ were found to contribute to less than 1\%
to all the selected events and are therefore neglected.

For all processes, the parton shower, multiple parton interactions, and hadronization
is simulated using \PYTHIA~8.230 with the CP5 tune~\cite{Sirunyan:2019dfx}.
The matrix element calculation and parton shower simulation are matched using the
FxFx~\cite{Frederix:2012ps} and MLM~\cite{Alwall:2007fs} algorithms for the processes
simulated with \MGvATNLO at NLO and LO, respectively.
All simulated samples use the NNPDF3.1~\cite{NNPDF:2017mvq} parton distribution functions~(PDFs) at next-to-NLO accuracy.

The interaction of particles with and response of the detector are simulated with the \GEANTfour package~\cite{Agostinelli:2002hh, Allison:2006ve}.
In addition, \pp inelastic collisions simulated with \PYTHIA are mixed into each event
to model the effect from additional collisions within the same or adjacent bunch crossing~(pileup).

\section{Event reconstruction}
\label{sec:reconstruction}

The particle-flow~(PF) algorithm~\cite{Sirunyan:2017ulk} aims to reconstruct and identify
each individual particle in an event, using an optimised combination of information from the
various elements of the CMS detector.
The primary vertex is taken to be the vertex corresponding to the hardest scattering in the event,
evaluated using tracking information alone, as described in Section 9.4.1 of Ref.~\cite{CMS-TDR-15-02}.
The missing transverse momentum vector \ptvecmiss is computed as the negative vector sum of the transverse momenta
of all the PF candidates in an event, and its magnitude is denoted as \ptmiss~\cite{Sirunyan:2019kia}.
The \ptvecmiss is modified to account for corrections to the energy scale of the reconstructed jets in the event.

The energy of photons is obtained from the ECAL measurement~\cite{CMS:2020uim}.
The energy of electrons is determined from a combination of the track momentum at the main interaction vertex,
the corresponding ECAL cluster energy, and the energy sum of all bremsstrahlung photons attached to the track.
Muons are reconstructed from tracks in the inner tracker that extend to the muon system.
From the curvature of the track, their momentum can be precisely determined~\cite{Sirunyan:2018fpa}.
Both muons and electrons must pass tight quality criteria developed by the CMS Collaboration
to ensure a proper reconstruction~\cite{Sirunyan:2018fpa, CMS:2020uim}.

The energy of charged hadrons is determined from a combination of the track momentum and the corresponding ECAL and HCAL energy deposits,
corrected for the response function of the calorimeters to hadronic showers.
Finally, the energy of neutral hadrons is obtained from energy measurements in ECAL and HCAL~\cite{Sirunyan:2017ulk}.

Electrons are further required to have $\pt>10\GeV$ and $\abs{\eta}<2.5$, excluding $1.44 < \abs{\eta} < 1.57$ that corresponds to the ECAL transition region.
Muons are selected with $\abs{\eta}<2.4$ and the same \pt requirement as that for the electrons.
In order to be considered for the analysis, electrons and muons must pass additional quality criteria~\cite{CMS:2020uim,Sirunyan:2018fpa}, that
concern the number of reconstructed hits in the tracker and/or muon
system, their compatibility with originating from the primary vertex, the quality of
charge reconstruction, and isolation from other activities in the detector.
In addition, we make use of a multivariate analysis that is trained to distinguish prompt leptons
from those that originate from heavy-flavour hadron decays and/or misidentified jets~\cite{CMS:2023ftu,CMS:2023dsu}, referred to as nonprompt leptons.
Following the definition in Ref.~\cite{CMS:2023ftu}, two identification working points~(WPs), `loose' and `tight', are used.
Tight leptons are used in the selection of events in the signal region, while the estimation of nonprompt backgrounds is based
on a selection of leptons passing the loose but failing the tight WP in dedicated control regions in data.
The efficiency and misidentification rate for selecting tight muons with $\pt>25\GeV$ ($10<\pt<25\GeV$)
were estimated in \ttbar simulation to be 96 (75)\% and 1 (0.9)\%~\cite{CMS:2023ftu}, respectively.
For electrons with $\pt>25\GeV$ ($10<\pt<25\GeV$), the selected WP corresponds to a selection efficiency and misidentification rate
of 93 (64)\% and 0.6 (0.4)\%~\cite{CMS:2023ftu}.

For each event, hadronic jets are clustered from these reconstructed PF particles
using the infrared and collinear safe \antikt algorithm~\cite{Cacciari:2008gp, Cacciari:2011ma} with a distance parameter of 0.4.
Jet momentum is determined as the vectorial sum of all particle momenta in the jet,
and is found from simulation to be, on average, within 5 to 10\% of the true momentum over the whole \pt spectrum and detector acceptance.
For jets to be considered in this analysis, they are required to have $\pt>30\GeV$ and be within $\abs{\eta}<2.4$.
Pileup can contribute additional tracks and calorimetric energy depositions to the jet momentum.
To mitigate this effect, charged particles identified to be originating from pileup vertices are discarded and an offset correction is applied to correct for remaining contributions.
Jet energy corrections~\cite{Khachatryan:2016kdb} are derived from simulation to bring the measured response of jets to that of particle level jets on average.
The energy scale corrections are propagated to \ptmiss.

Jets initiated by the hadronization of \PQb quarks~(\PQb jets) are identified using the \textsc{DeepJet} algorithm~\cite{CMS-DP-2018-033, Bols:2020bkb}.
The WP chosen for this analysis corresponds to a misidentification rate of 0.1\% for light-flavour quark and gluon jets,
and an efficiency of 68\% for \PQb jets.

\section{Event selection}
\label{sec:selection}

The data for this analysis are selected using a combination of electron and muon triggers.
With respect to the baseline selection detailed below, we obtain a trigger efficiency of more than 99\%,
using a combination of triggers that require single or multiple leptons with \pt thresholds between 5 and 40\GeV.
Further selection criteria are defined offline in order to select regions that contain a mostly pure sample of the target processes \ttZ, \WZ, and \ZZ.
For each process, a dedicated SR is defined, resulting in three regions \SRttZ, \SRWZ, and \SRZZ.
The nonprompt lepton contributions in \SRttZ and \SRWZ are derived from CRs
that are labelled \CRttZ and \CRWZ.
In \SRZZ, the contribution from nonprompt leptons is negligible and no corresponding CR is introduced.

All SRs and CRs are based on a common baseline selection.
This baseline selection requires at least three leptons (electrons or muons) that pass the loose WP,
with $\pt > 40$, 20, and 10\GeV.
We do not explicitly veto electrons or muons from leptonic tau lepton decays.
To remove leptonic decays of low-mass resonances, events are selected where
all pairwise combinations of leptons have an invariant mass above 12\GeV.
Furthermore, at least one lepton pair of same flavour and opposite electric charge is required to have
an invariant mass compatible with the \PZ boson mass of 91\GeV~\cite{PDG2024} ($81 < \mll < 101\GeV$).
If multiple pairs satisfy these criteria, the pair closest to the \PZ boson mass
is considered a \PZ boson candidate.
For events with four leptons, a second \PZ boson candidate, with similar requirements, is constructed from the
remaining two leptons after the pair of leptons closest to the \PZ boson mass is
removed from the list.

\begin{figure*}[htb!]
  \centering
    \includegraphics[width=0.49\textwidth]{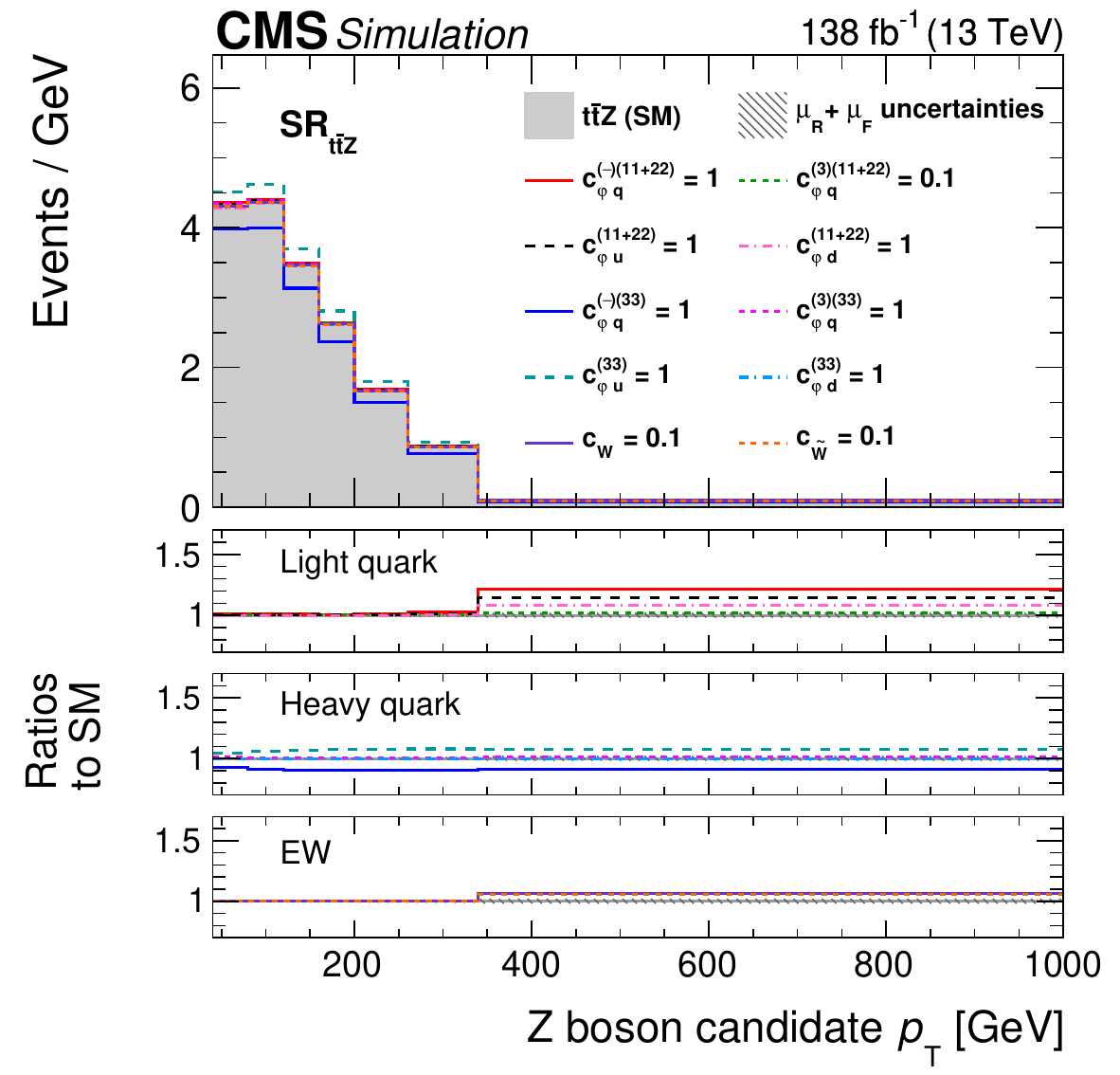}
    \includegraphics[width=0.49\textwidth]{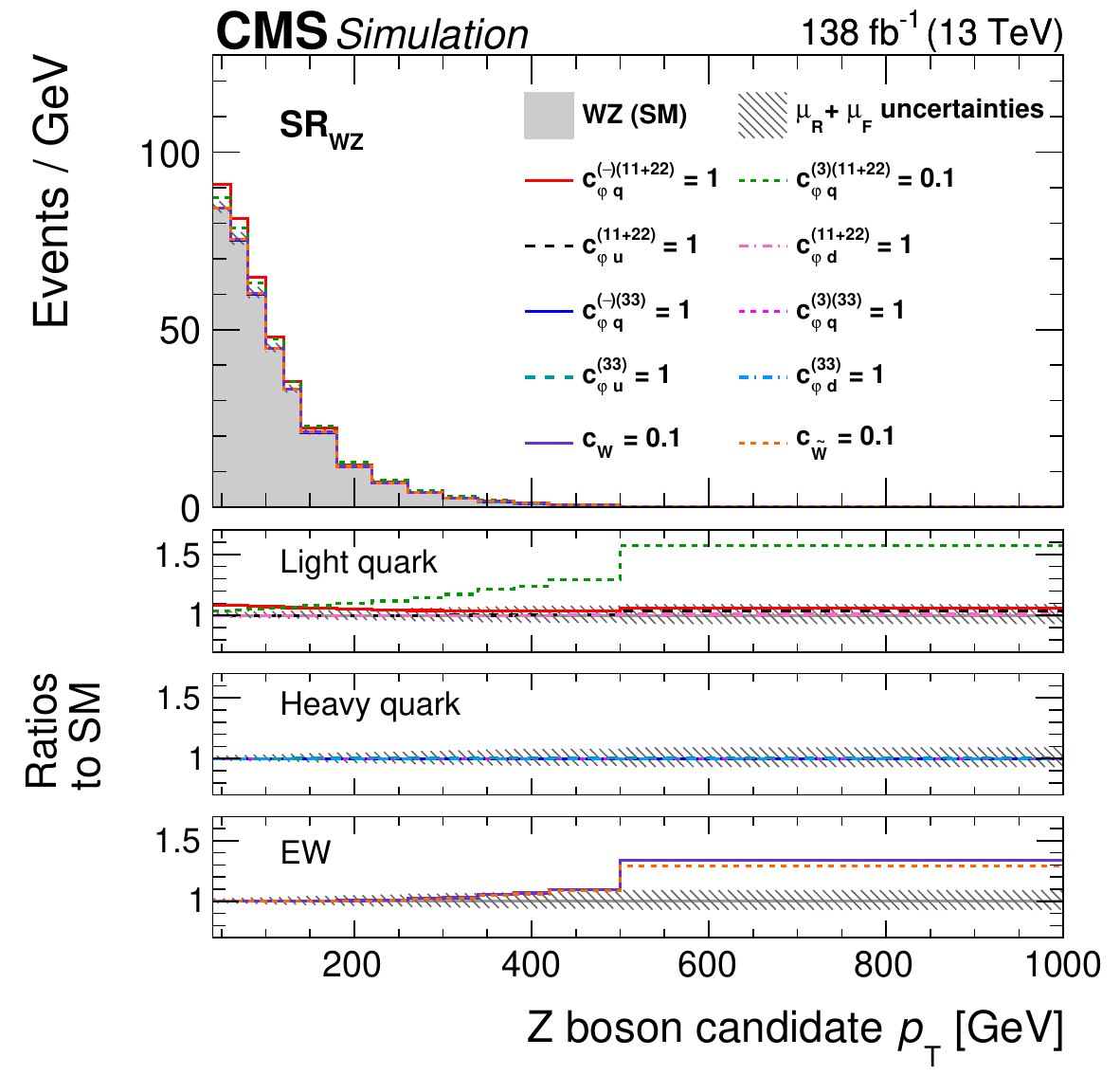}
    \includegraphics[width=0.49\textwidth]{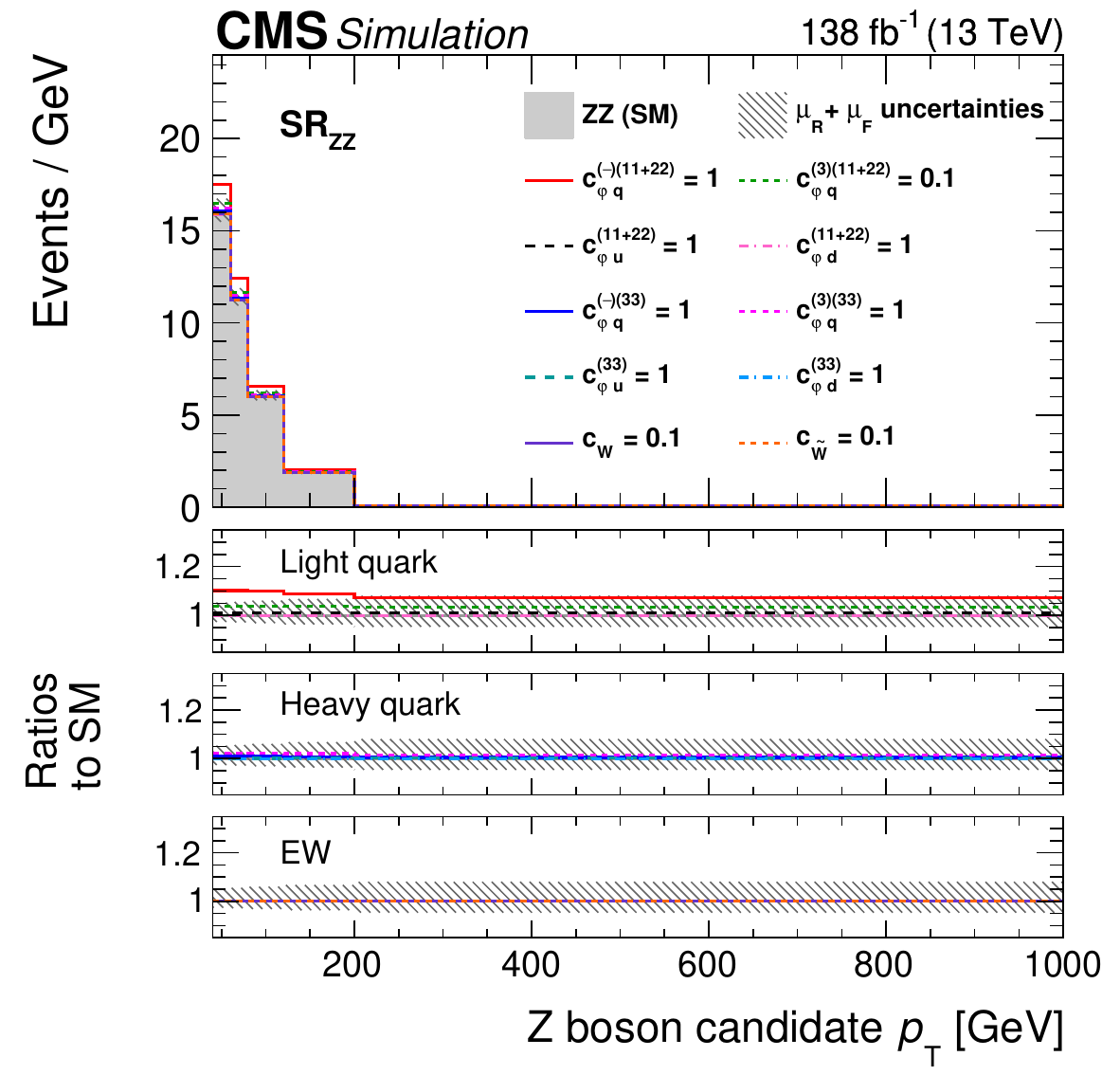}
    \caption{Distributions of the \PZ boson \pt in the three signal regions of this analysis.
    Shown are \SRttZ~(upper left), \SRWZ~(upper right), and \SRZZ~(lower).
    In each region, the target process (\ttZ, \WZ, or \ZZ) is shown at the SM point~(coloured areas)
    and various EFT hypotheses~(lines).
    The hashed band includes only uncertainties in the renormalisation and factorisation scales~($\mu_{\text{R}}$ and $\mu_{\text{F}}$).
    The upper, middle, and lower ratio panels show the ratio of EFT hypotheses for light-quark,
    heavy-quark, and EW boson couplings, respectively.
    The bin content is divided by the bin width.
    \label{fig:EFT_SR}}
\end{figure*}

The signal region \SRttZ additionally requires events
to contain exactly three tight leptons without additional loose leptons.
A large fraction of nonprompt leptons are rejected as a result of this selection and the remaining ones are estimated from data.
To reduce the \WZ contribution, at least one jet must be \PQb tagged.
The contribution of \tZq production is reduced by requiring at least three jets.
From simulation, we expect the yield to be about 60\% \ttZ events with
a subleading contribution of up to 10\% from \WZ production.
In \SRWZ, events are required to pass the baseline selection and
to contain exactly three leptons passing the tight WP.
Events with additional loose leptons are vetoed.
To enhance leptonic decays of the \PW boson, events with $\ptmiss>60\GeV$ are selected.
In order to construct a region orthogonal to \SRttZ and suppress processes
with top quarks, events without any \PQb-tagged jets are selected.
In \SRWZ, 87\% of the events originate from \WZ production.
The region \SRZZ is defined by the baseline selection,
requiring four leptons with tight identification criteria.
Events must contain two \PZ boson candidates following the requirements detailed earlier.
The region is very pure in \ZZ production (99\%) and has a negligible background contribution.
The effects of the EFT operators probed in this analysis are displayed
in each of the SRs in Fig.~\ref{fig:EFT_SR},
where the target process in each of the three SRs is shown at the SM point
and compared to selected EFT hypotheses.
It is clearly visible that the \ttZ process is sensitive to the third-generation couplings,
while \WZ and \ZZ can constrain the couplings of light quarks to EW bosons.
Furthermore, the \ttZ process adds to the sensitivity of the first and second generations
since the \PZ boson can be radiated from initial-state light quarks.
The radiation from initial-state quarks is enhanced at large \PZ boson \pt, which
increases the sensitivity to light-quark couplings and reduces it for heavy-quark
couplings in the \ttZ process at high \pt.

\begin{figure*}[tb]
  \centering
    \includegraphics[width=0.49\textwidth]{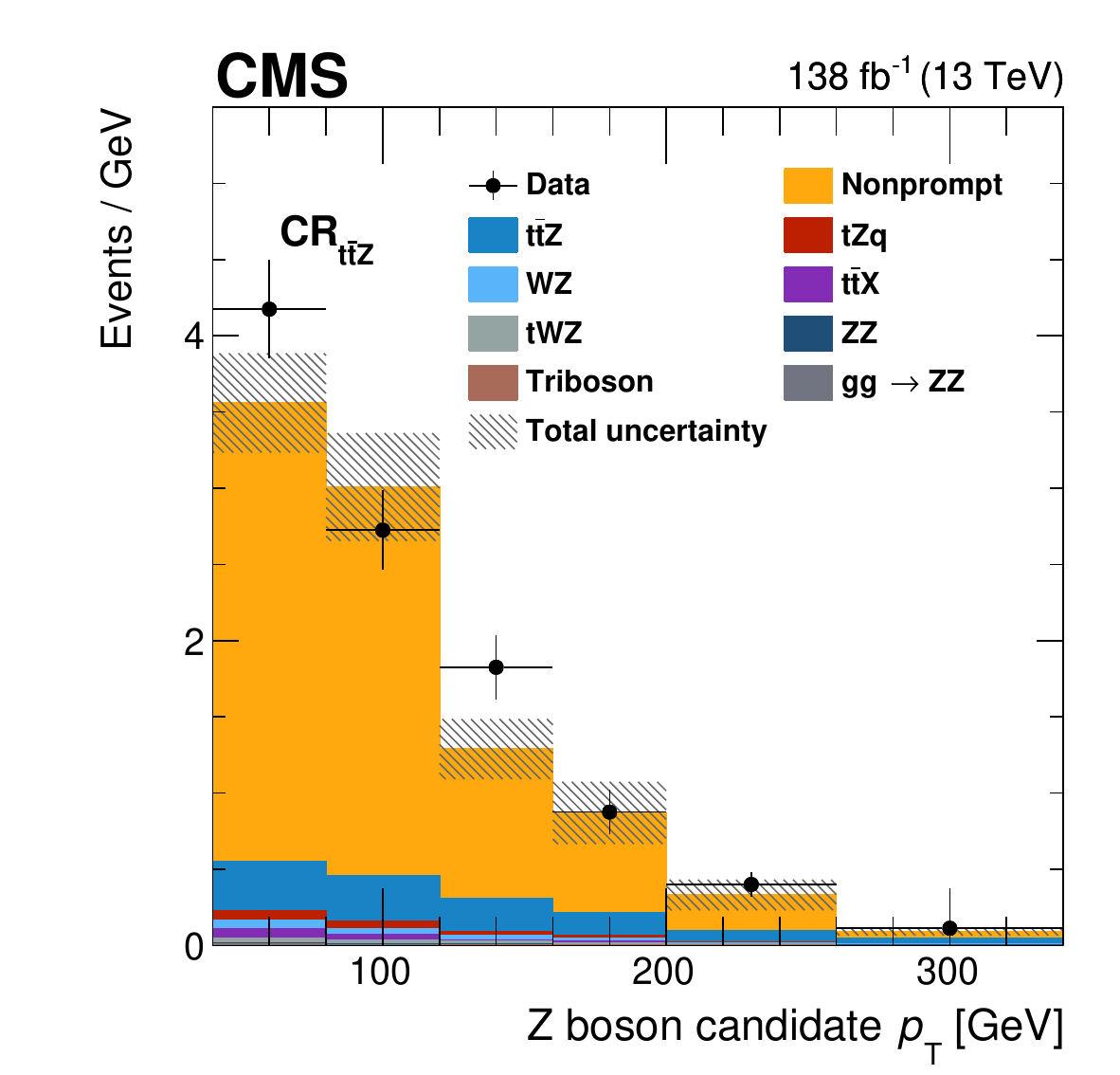}
    \includegraphics[width=0.49\textwidth]{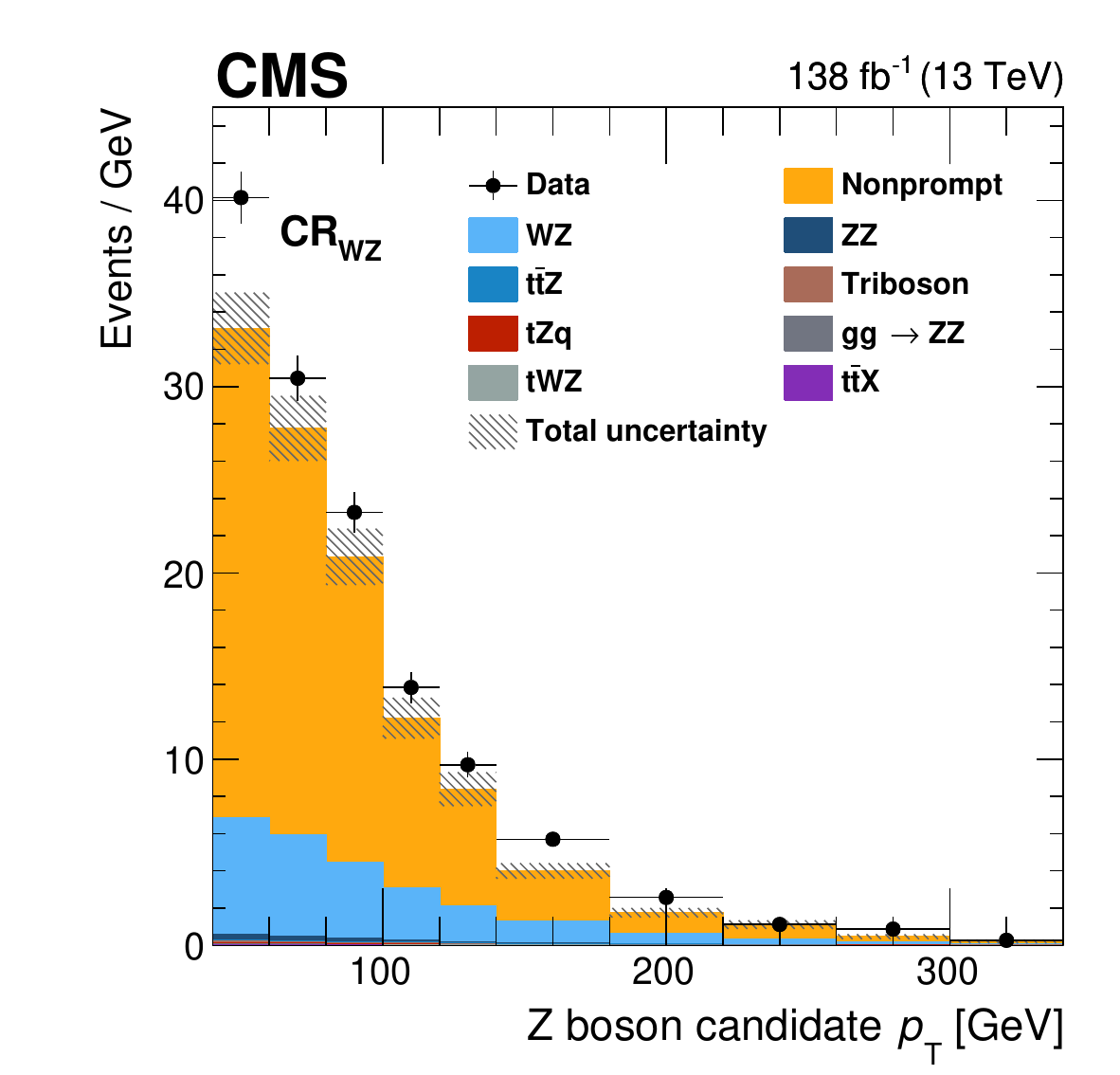}
    \caption{Distributions of the \PZ boson \pt in the control regions of this analysis.
    Shown are \CRttZ~(left) and \CRWZ~(right).
    Predictions are all obtained from simulation and are displayed as coloured areas.
    The hashed area shows the statistical uncertainty in the prediction.
    Data are displayed as markers, where the vertical bars represent the statistical
    uncertainty.
    The bin content is divided by the bin width.
    \label{fig:CRs}}
\end{figure*}

The control regions \CRttZ and \CRWZ have the same criteria as the corresponding SRs with the exception that at least
one of the selected leptons must fail the tight WP of the lepton identification.
Figure~\ref{fig:CRs} shows the \PZ boson \pt distributions in the two CRs for data and simulation.
They are enriched in contributions from nonprompt leptons and allow the
estimation of the nonprompt contribution in the signal regions \SRttZ and \SRWZ as further discussed in the following section.
The nonprompt contribution shown for illustration in Fig.~\ref{fig:CRs} is obtained from simulated events.

\section{Background estimation}
\label{sec:nonprompt}
The processes targeted in this analysis are rare in comparison with other processes involving final-state leptons at the LHC.
Thus, despite large lepton identification efficiencies and small misidentification rates,
processes with nonprompt leptons significantly contribute to the selected SRs.
The origin of nonprompt leptons is manifold.
In the SRs of this analysis, nonprompt leptons mostly occur in Drell--Yan or \ttbar production, where the third lepton is nonprompt.
We distinguish nonprompt electrons, which tend to be misidentified jets in Drell--Yan processes, and
nonprompt muons, which mostly originate from heavy-flavour decays in the parton shower, \eg in \ttbar events.

Since nonprompt leptons are difficult to model in simulation,
their contribution is estimated from data~\cite{CMS:2018fdh,CMS:2020mpn,CMS:2023xyc,CMS:2023ftu} by
using \CRttZ and \CRWZ.
To estimate the contribution in the SRs, a transfer factor, $w$, is applied
to the nonprompt contribution measured in the CRs.
It is a function of the misidentification rate, \ie the probability that a nonprompt lepton
that passes the loose identification also passes the tight WP.
The transfer factor is
\begin{equation}
w = (-1)^{\Nlep-1} \prod^{\Nlep}_{i} \frac{f_i}{1-f_i},
\end{equation}
where $f_i$ is the misidentification rate of the $i$-th lepton that does not fulfil the tight identification
and \Nlep is the total number of leptons that fail the tight identification.
The sign flip eliminates double counting~\cite{CMS:2018fdh} of the probability in events where multiple
leptons do not pass the tight identification.

\begin{figure*}[tb]
  \centering
    \includegraphics[width=0.98\textwidth]{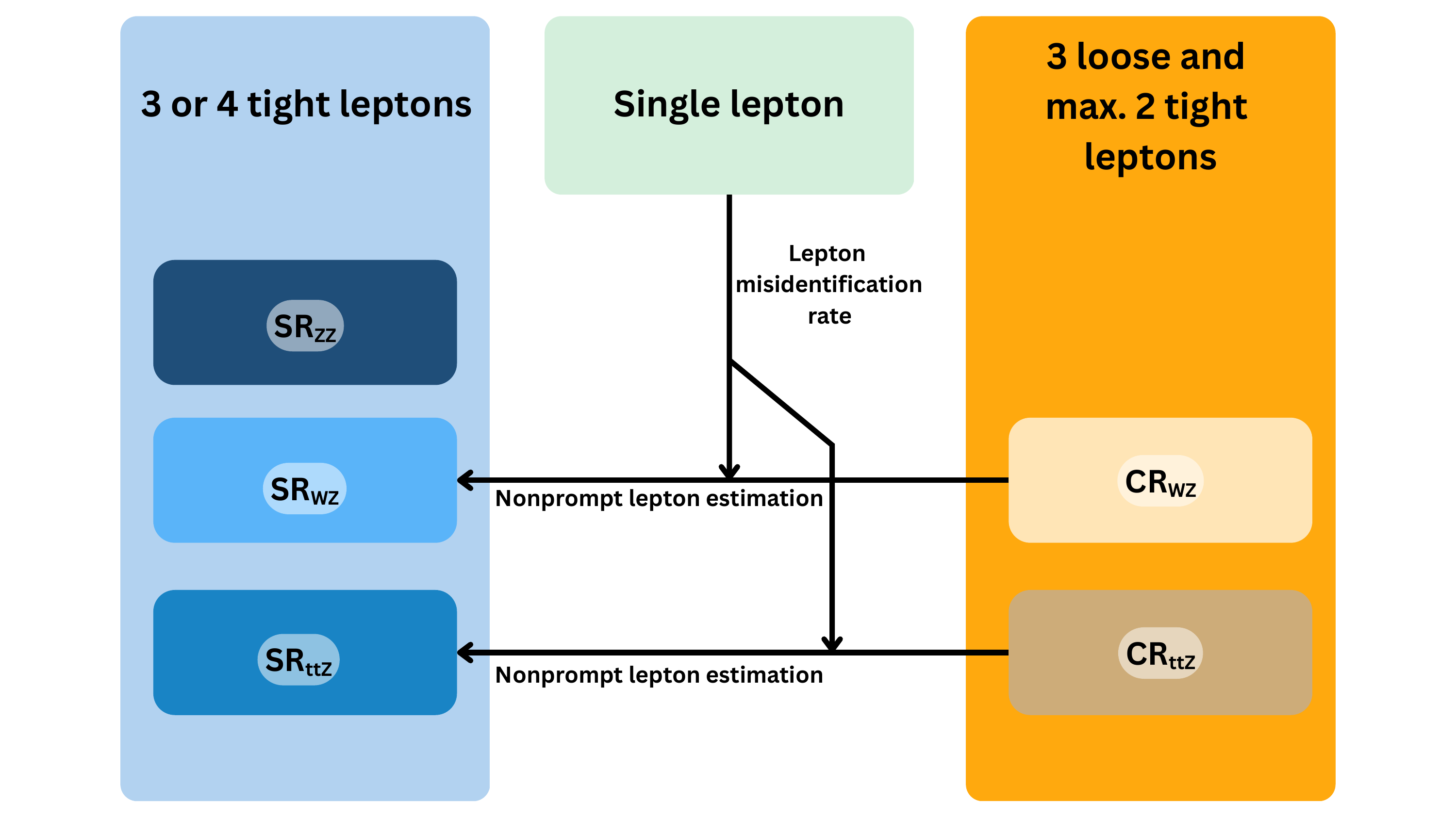}
    \caption{Schematic representation of the SRs and CRs used in this analysis.
    The application of the estimated nonprompt lepton background from the CRs into the SRs is illustrated with arrows.
    \label{fig:regions}}
\end{figure*}

The misidentification rate $f$ is measured in additional single-lepton CRs.
In these CRs, events are selected with single electron or single muon triggers
with muon (electron) \pt thresholds as low as 3 (8)\GeV;
we require that events have exactly one lepton with $\pt > 10\GeV$ and $\abs{\eta}<2.4$ (2.5)
that passes the loose WP,
at least one jet with an angular separation $\Delta R > 0.7$ from the lepton, and $\ptmiss < 20\GeV$.
The selection criteria are chosen such that a rich sample of nonprompt leptons is selected.
The misidentification rate $f$ is measured in data after subtracting the prompt contribution estimated from simulation and
differentially in lepton flavour (electrons and muons), \pt, $\eta$, and year of data taking.
It has values between 0.15 and 0.8.
A schematic representation of all SRs and CRs is displayed in Fig.~\ref{fig:regions},
together with the strategy for estimating the nonprompt lepton background
in the SRs.

\begin{figure*}[tb]
  \centering
    \includegraphics[width=0.49\textwidth]{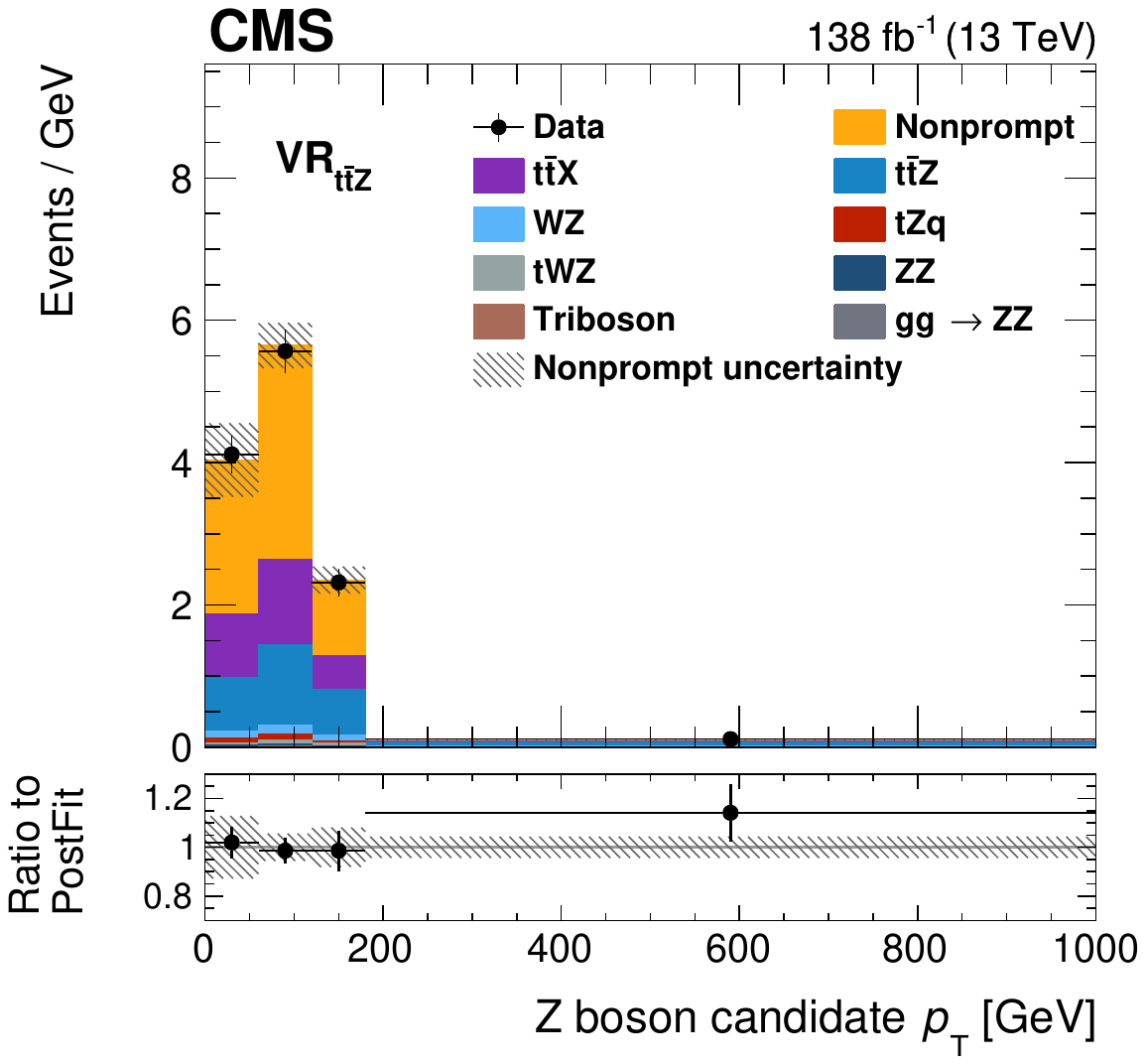}
    \includegraphics[width=0.49\textwidth]{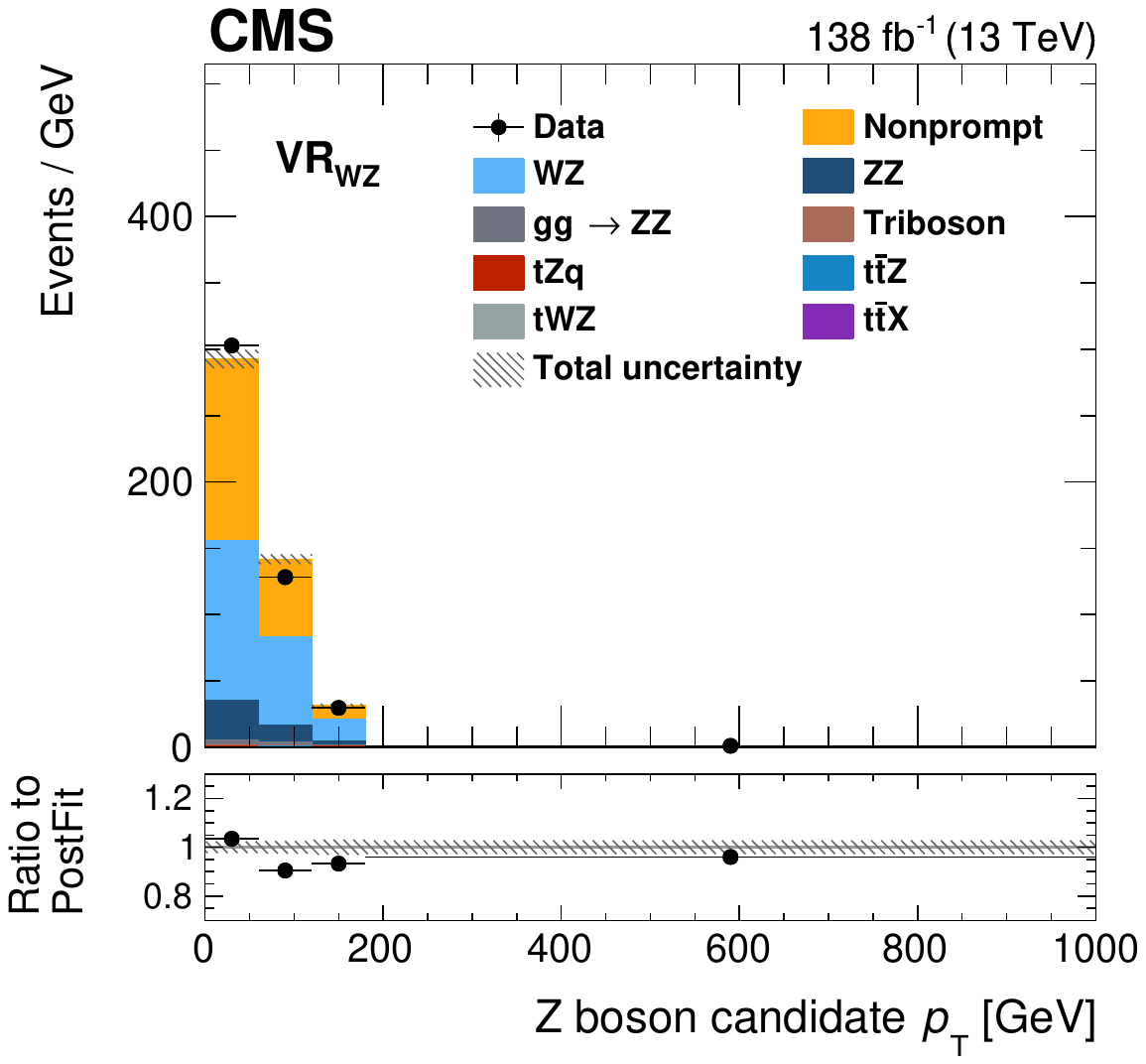}
    \caption{Distributions of the \PZ boson \pt in the two validation regions \VRttZ~(left)
    and \VRWZ~(right) after a fit including only uncertainties in the nonprompt lepton estimation.
    The data~(markers) are compared to the prediction from simulation and
    the data-driven estimate of nonprompt leptons~(coloured areas).
    The lower panel displays the ratio to the predictions after the fit.
    The hashed area displays the uncertainties in the estimation of the nonprompt lepton background after the fit.
    The bin content is divided by the bin width.
    \label{fig:validationRegions}}
\end{figure*}

In addition to the statistical uncertainty in the measurement of the misidentification rates,
a modelling uncertainty is assigned to account for potential nonclosure,
which is estimated in data and simulation.
In data, the background estimation is tested in validation regions~(VRs) orthogonal to \SRttZ and \SRWZ,
where all selection criteria remain, but no \PZ boson candidate in the mass range 81--101\GeV can be formed from two opposite-sign leptons of same flavour~(\VRttZ)
or the missing transverse momentum is below 60\GeV~(\VRWZ).
In the VRs, the contribution of nonprompt leptons is enhanced with respect to the SRs,
and the data agree with the estimated event yields after a fit that only includes the 
uncertainties related to the background estimation,
as shown in Fig.~\ref{fig:validationRegions}.
For the estimation of the nonprompt lepton contribution in the VRs, dedicated CRs are constructed analogously to those used for the estimation in the SRs.
Along with the \PZ candidate \pt, other observables such as the \pt distributions of the three
selected leptons were tested as well.
In addition to the test in data, we fit the estimated contribution of nonprompt leptons
to the true contribution in simulation and verify that the uncertainty model is flexible
enough to fit a residual nonclosure.

In total, the nonclosure uncertainty consists of 15 components, divided into
three lepton flavour categories and five correlation patterns.
For the flavour categories, we identify if only muons, only electrons, or a mixture
of both do not satisfy the tight lepton identification criteria.
The correlation patterns assign each of the data-taking periods an uncorrelated
part and one that correlates the uncertainty between all periods.
Each uncertainty amounts to a normalisation uncertainty of 25\%.
We observe closure in all tests when considering only the uncertainties corresponding to the nonprompt-lepton background estimation.

Contributions from photon conversions through interactions with the detector material
are drastically reduced by the lepton identification criteria.
Studies with simulated \ttgamma events showed that this contribution is negligible in all regions of the analysis.

All processes with prompt leptons---\ttZ, \WZ, \ZZ, \tZq, \tWZ,
\ttX, or the production of three EW bosons---are
estimated from simulation, as described in Section~\ref{sec:dataMC}.

\section{Systematic uncertainties}
\label{sec:uncertainties}
Systematic uncertainties are split into experimental and theory components,
depending on their origin.
Each uncertainty is implemented as one or multiple nuisance parameters in the profiled
likelihood fit and is evaluated for both the SM and the EFT predictions.
For most extracted WCs, the uncertainties in the production cross section of the \ttZ, \WZ, and \ZZ
processes have the largest impact among the systematic uncertainties.

\subsection{Experimental uncertainties}
Experimental uncertainties arise from the calibration of physics objects that are
used in the analysis, from efficiency corrections in the simulation, and from the estimation of background processes.
The uncertainties are implemented in this analysis by propagating shifts of ${\pm}1$ standard deviations to the distributions used in the fit.

The uncertainty in the lepton identification~\cite{Sirunyan:2018fpa, CMS:2020uim} efficiency is measured
as a function of the lepton flavour, \pt, and $\eta$.
Based on whether the uncertainties have statistical or systematic origin, they are
split into uncorrelated and correlated parts for the four data-taking periods,
since the year 2016 is split into two periods.
This uncertainty is modelled with a total of 10 nuisance parameters in the profiled
likelihood fit discussed below.
An additional parameter models the uncertainty in the electron reconstruction efficiency,
which is treated as correlated for data-taking periods.

The uncertainty in the correction of the \PQb tagging~\cite{Sirunyan:2017ezt} efficiency is a function of jet \pt, $\eta$,
\PQb tagging WP, as well as flavour.
The uncertainty is split in correlated and uncorrelated parts across the data-taking periods
and is modelled with 10 nuisance parameters.

The pileup profile in the simulation is weighted to match the
pileup profile observed in data~\cite{CMS:2020ebo}.
Observables sensitive to pileup are used to estimate the effective value for the
inelastic \pp cross section of $69.2\mb \pm 4.6\%$,
where the uncertainty accounts for remaining differences between data and simulation.
In this analysis, the event weight is varied within this uncertainty.
The uncertainty is modelled with a single parameter.

The efficiency of the combination of triggers~\cite{Khachatryan:2016bia} used in this analysis is measured
using a sample of events collected with an independent trigger to be larger than 99\% in the selected phase space for data and simulation.
We do not assign any correction of the trigger efficiency in simulation but consider a 1\% uncertainty
that represents the statistical uncertainty in the efficiency measurement in data for each data-taking period,
resulting in four nuisance parameters.
A gradual shift in the timing of the ECAL and muon systems causes inefficiencies of the first-level trigger
system that is not included in the simulation~\cite{CMS:2020cmk}
and not accounted for by the aforementioned efficiency measurements.
With auxiliary measurements of these inefficiencies in data, the effect is modelled in simulation.
A corresponding systematic uncertainty is modelled with a single nuisance parameter.

In the estimation of the nonprompt background, the precision of the measured misidentification rate
is limited by the number of selected events in the single lepton CRs where the lepton misidentification rate is measured.
This statistical uncertainty is split by lepton flavour and data-taking period and
propagated to the estimation of the nonprompt background
in the SRs.
In addition, various normalisation uncertainties are assigned to cover the potential nonclosure
of this method.
We split into an uncorrelated and a correlated part across the data-taking periods as well as
split the uncertainty according to the misidentified-lepton flavour.
As detailed in Section~\ref{sec:nonprompt}, this leads to a total of 15 nuisance parameters.

The integrated luminosity is measured with a precision of 1.2, 2.3, and 2.5\%
in the data-taking periods of 2016, 2017, and 2018~\cite{lumi16, lumi17, lumi18}.
The luminosity uncertainties are partly correlated between the three years.
This results in uncorrelated uncertainties of 1.0, 2.0, and 1.5\% for the three years.
A correlated part among all data-taking years contributes as 0.6, 0.9, and 2.0\% for
2016, 2017, and 2018.
Lastly, correlations between 2017 and 2018 amount to uncertainties of 0.6 and 0.2\%.
This results in a total of five nuisance parameters.

The jet four-momenta are varied within the uncertainties in the
jet energy scale, considering various uncertainty sources split between detector regions and
data-taking periods~\cite{Khachatryan:2016kdb}.
Also, the jet four-momenta are varied within the uncertainties in the jet energy
resolution smearing.
This uncertainty is considered uncorrelated for all data-taking periods such that
four nuisance parameters are introduced.
All these variations are also propagated to \ptmiss.

The missing transverse momentum is corrected for PF candidates that are
not clustered into jets.
We vary \ptmiss within the corresponding uncertainty using four parameters for uncorrelated
effects between data-taking periods.

\subsection{Theory uncertainties}
Various uncertainties are associated with our limited knowledge in the modelling of
all simulated processes relevant to this analysis.
This includes uncertainties that only affect the normalisation and those that
change the shape and acceptance.

The factorisation and renormalisation scales are varied independently by factors of 0.5 and 2
and we construct nuisance parameters for each process.
Here, only differences in the shape and acceptance of the distributions and not in the normalisation are considered.
In order to test the assumption of uncorrelated uncertainties,
the impact of correlations was studied in the \WZ and \ZZ processes by performing the profiled likelihood fit with fully correlated and
uncorrelated setups. The differences in the resulting likelihoods was found to be negligible.
Thus, per process, one nuisance parameter is introduced to model the uncertainty in the factorisation
scale and another parameter for the renormalisation scales, which amount to a total of 14
parameters in the likelihood fit.

Uncertainties in the proton PDFs are estimated by using NNPDF replicas~\cite{Ball:2014uwa,NNPDF:2017mvq},
resulting in 100 variations.
For each variation a nuisance parameter is constructed in the fit, correlating all processes.
Similar to scale variations, only shape and acceptance effects from the event selection are considered.
Normalisation effects are absorbed into uncertainties in the production cross section
described below.

Parton shower uncertainties are estimated by varying the energy scales of the
initial-state radiation~(ISR) and final-state radiation~(FSR) individually
by factors of 0.5 and 2.
Similar to scale variations, only shape and acceptance effects are considered.
The ISR uncertainties are split into nuisance parameters for each process, whereas
the FSR uncertainty is correlated between processes.
This results in a total of eight nuisance parameters for parton shower uncertainties.

We consider production-rate uncertainties arising from the limited accuracy of the computed process cross sections, 
uncertainties in dedicated measurements, and the small number of simulated events that pass the analysis selection.
These uncertainties only affect the normalisation of a process and are
correlated among all years. A 20\% uncertainty is assigned to \tWZ,
triboson production, as well as \ttX processes, which include
\ttW~\cite{CMS:2022tkv}, \ttH~\cite{Kulesza:2020nfh}, $\PQt\PAQt\PW\PW$, $\PQt\PAQt\PW\PZ$, $\PQt\PAQt\PZ\PZ$, and \tttt.
Although recent measurements of the inclusive \ttW cross section showed a much larger value compared to the
SM prediction, the \ttW contribution in this analysis is so small that enhancing the corresponding cross section
or its uncertainty does not alter the results.
We assign uncertainties of 10~\cite{CMS:2021ugv} and 11\%~\cite{Kulesza:2020nfh} to the
production cross sections of the \tZq and \ttZ processes, respectively.
The diboson processes \WZ and \ZZ are assigned rate uncertainties of 5\%~\cite{CMS:2020gfh,Cascioli:2014yka}.
This results in a total of seven nuisance parameters.
For the \tZq and \tWZ production, additional 20\% rate uncertainties are assigned
accounting for the missing EFT effects in the simulations of these processes,
estimated from the EFT effects observed in Ref.~\cite{CMS:2020lrr}.
Since those processes only contribute a small amount in our selected phase space,
this uncertainty has a negligible effect on the extracted WCs.

The modelling of the number of associated jets in \WZ production in a similar phase space
was found to be not well described
in the observation of the \tttt production~\cite{CMS:2023ftu}.
The correction factor is applied as a function of the number of jets and reaches
from 0.85 for one jet to 2.69 for six jets.
The difference between not applying and
applying the correction factors measured in Ref.~\cite{CMS:2023ftu} is used as an uncertainty,
which is implemented as a single nuisance parameter.
However, the correlation with the sensitive observable used in this analysis, the
reconstructed \PZ boson \pt, is small such that this uncertainty is only
relevant in \SRttZ, where three jets are required.
Here the uncertainty only impacts the total \WZ event yield by about 20\% in \SRttZ.
With a signal injection test it is ensured that a varied modelling of jet multiplicity
in \WZ production does not bias the extraction of EFT effects.

In order to estimate the uncertainty in the modelling of \WZ plus heavy-flavour quarks,
an additional 20\% normalisation uncertainty is assigned for simulated \WZ events
that are produced in association with \PQc and \PQb quarks~\cite{CMS:2021ugv}.
Also this uncertainty is modelled with a single parameter in the fit.

For the processes \WZ and \ZZ we assign an uncertainty corresponding to the estimated
EW corrections to the production cross section obtained in Ref.~\cite{Grazzini:2019jkl}.
The uncertainty is implemented as a function of the true \PZ boson \pt in simulation.
Four nuisance parameters are implemented to model the additive and multiplicative corrections to \WZ and \ZZ production.

\section{Maximum likelihood fit and results}
\label{sec:results}
\begin{figure*}[tbp]
  \centering
    \includegraphics[width=0.49\textwidth]{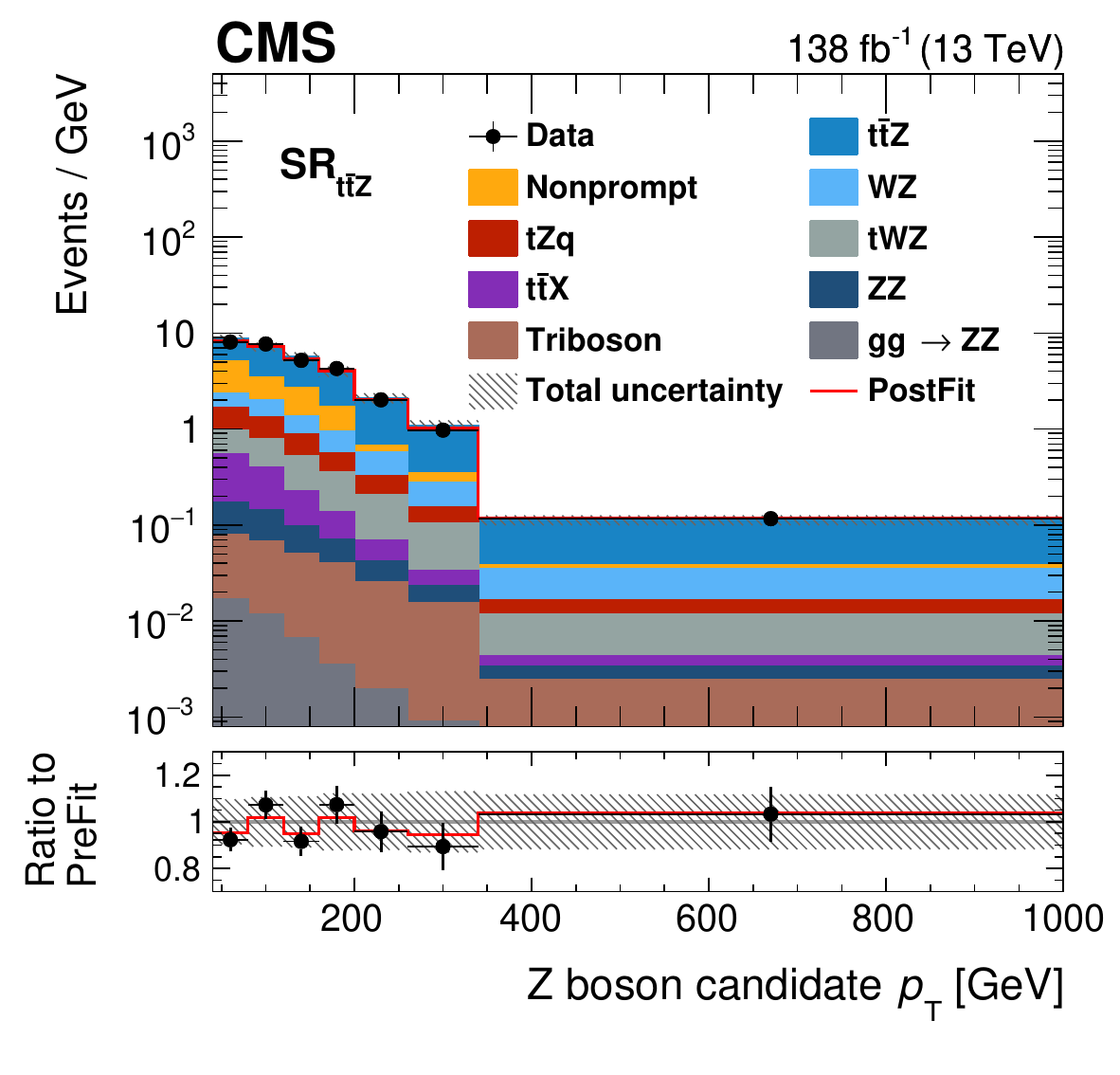}
    \includegraphics[width=0.49\textwidth]{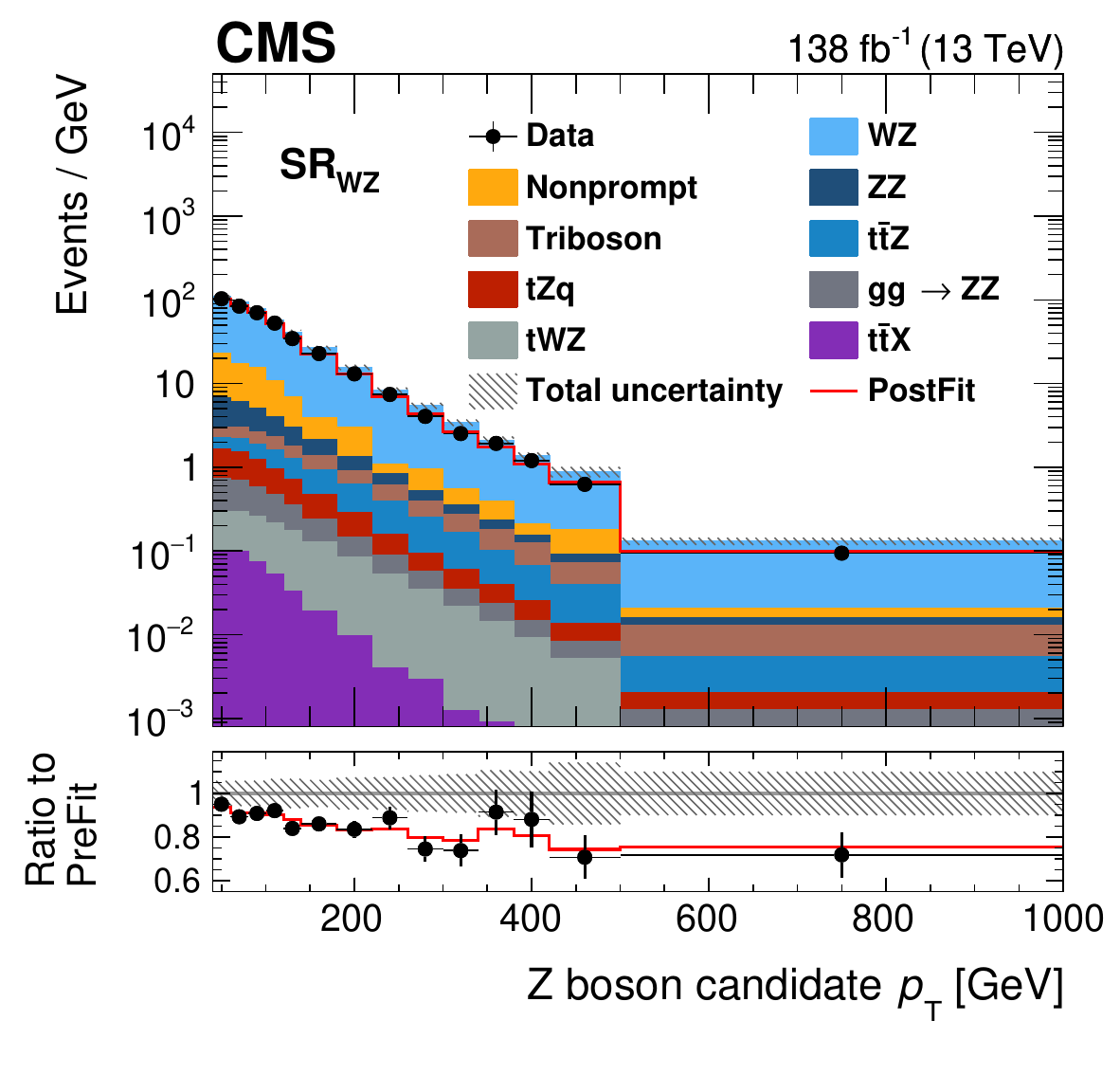}
    \includegraphics[width=0.49\textwidth]{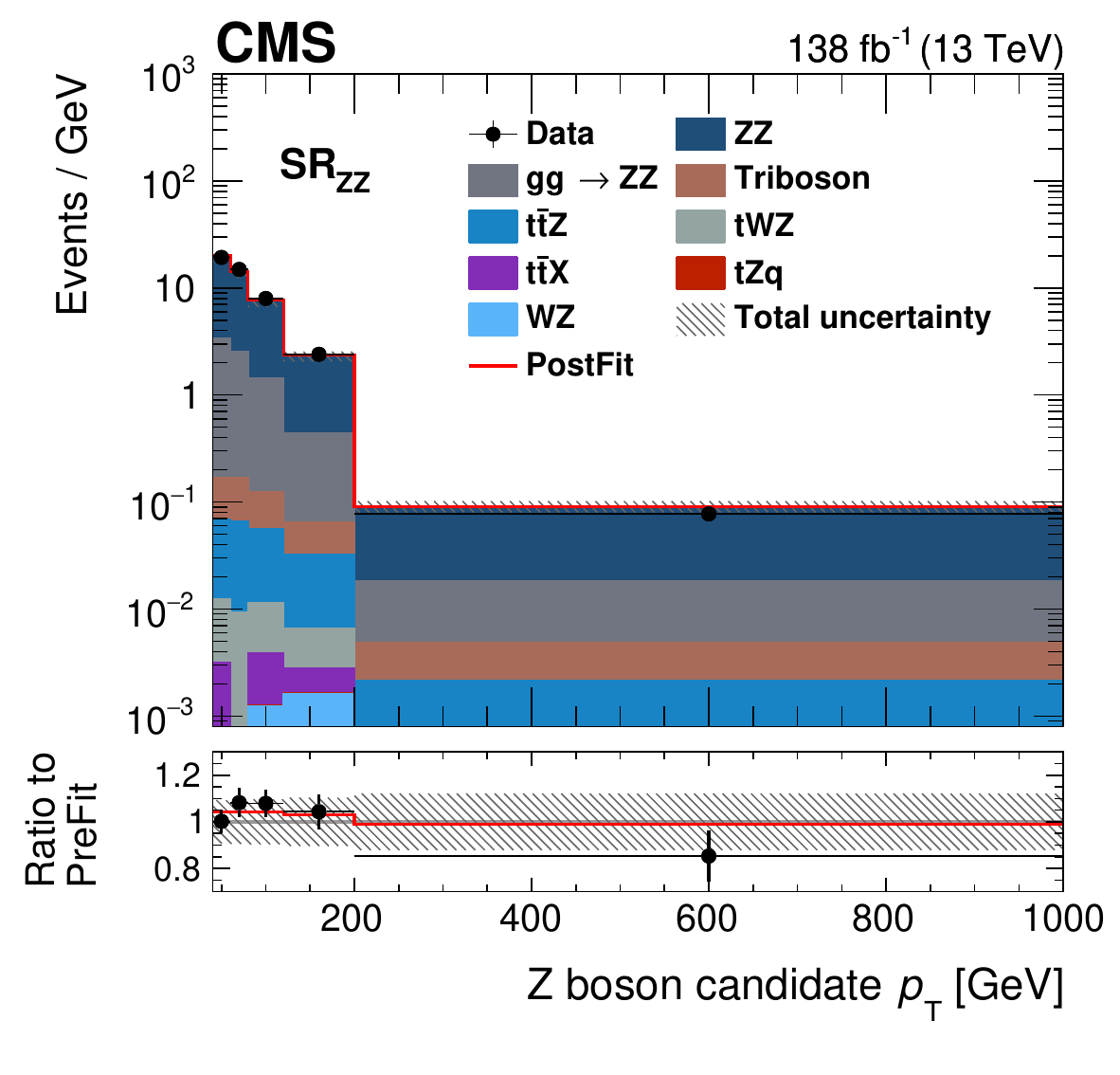}
    \caption{Distributions of the \PZ boson \pt in the three signal regions of this analysis.
    Shown are the \SRttZ~(upper left), \SRWZ~(upper right), and \SRZZ~(lower) regions.
    The data~(markers) are compared to the prediction from simulation and
    the data-driven estimate of nonprompt leptons~(coloured areas).
    The lower panel displays the ratio to the predictions before the fit.
    The hashed area displays the total uncertainties before the likelihood fit and the red line
    displays the best fit result in a setup with all EFT parameters.
    The bin content is divided by the bin width.
    \label{fig:signalRegionsPostFit}}
\end{figure*}

We extract the values of the WCs in a maximum likelihood fit
to data, using the CMS statistical analysis tool \textsc{Combine}~\cite{CMS:2024onh},
based on the \textsc{RooFit}~\cite{Verkerke:2003ir} and \textsc{RooStats}~\cite{Moneta:2010pm} frameworks.
In this setup, the WCs are the parameters of interest, 
and we estimate the effects of terms in the matrix element that depend linearly and quadratically on the WCs.
We simultaneously fit the \PZ boson \pt distributions in the regions \SRttZ, \SRWZ, and \SRZZ,
allowing for sensitivity to all operators, while other observables---such as angular distances between
leptons or \PZ boson candidates---do not result in a more sensitive probe.
The likelihood is built from Poisson probabilities to observe the per-bin event yields in data
given the signal and background predictions.
Sources of systematic uncertainties are modelled with nuisance parameters in the fit.
Shape uncertainties are implemented using a Gaussian distribution and
rate uncertainties are modelled with a log-normal distribution.
Statistical uncertainties related to the finite number of simulated events are modelled
with a single nuisance parameter per bin in the fit~(Barlow--Beeston lite)~\cite{Barlow:1993dm}.
Figure~\ref{fig:signalRegionsPostFit} shows the fitted distributions, including the best fit point.
The total uncertainty after the likelihood fit remains similar to the PreFit uncertainty, 
as the sensitivity to the WCs is largely driven by the uncertainties in the production rates, which are only weakly constrained by the fit.
We combine all years of data taking prior to the fit but also verified that we obtain
compatible results when fitting each data-taking period individually.

\begin{figure*}[tbp]
  \centering
    \includegraphics[width=0.7\textwidth]{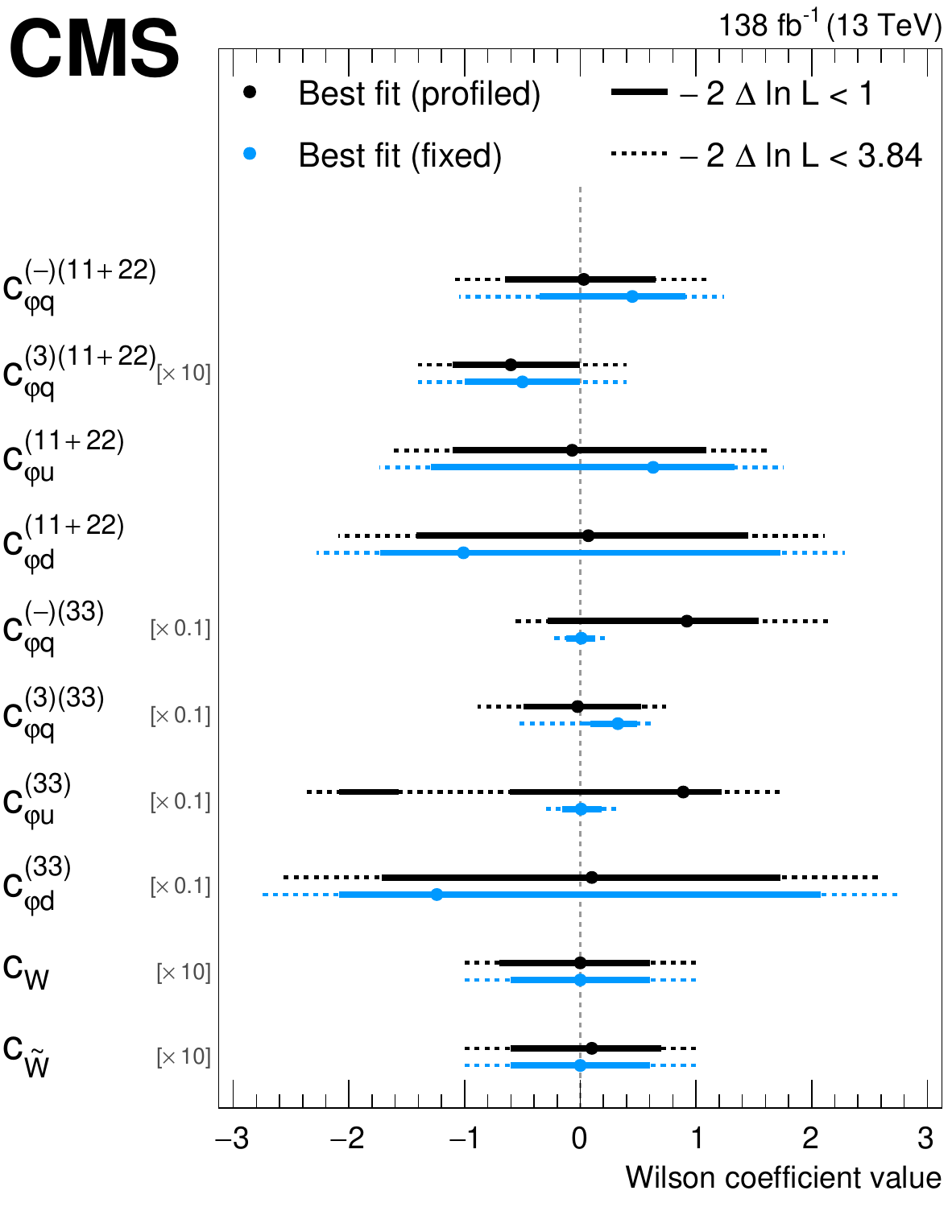}
    \caption{Summary of the limits obtained for the Wilson coefficients
    \cHqMinusLight, \cHqTripletLight, \cHuLight, \cHdLight, \cHqMinusThird, \cHqTripletThird, \cHuThird, \cHdThird, \cW, and \cWtil.
    Shown are the best fit points and limits
    for scans where other Wilson coefficients are fixed to zero~(`fixed') or are allowed to float~(`profiled').
    The points where the difference $-2\Delta\ln L$ with respect to the best fit increases by 1 and 3.84 are shown as horizontal error bars.
    These points correspond to the 68 and 95\% \CL limits in the asymptotic approximation.
    For each Wilson coefficient value, the EFT energy scale is assumed to be $\Lambda=1\TeV$.
    For better visibility, the heavy quark couplings are multiplied by a factor of 0.1 and 
    \cHqTripletLight, \cW, as well as \cWtil are multiplied by 10.
    \label{fig:summary}}
\end{figure*}

\begin{figure*}[tbp]
  \centering
    \includegraphics[width=0.49\textwidth]{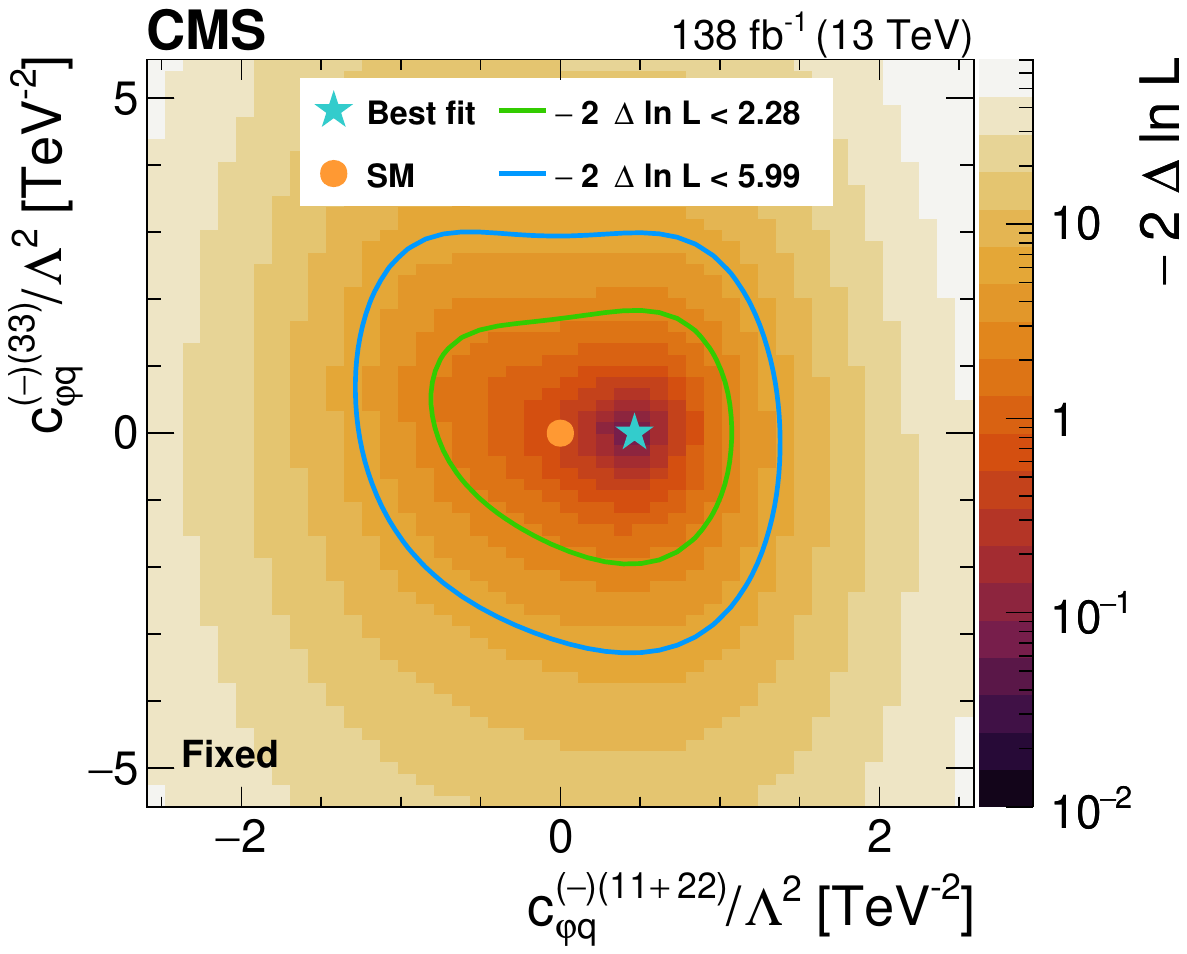}
    \includegraphics[width=0.49\textwidth]{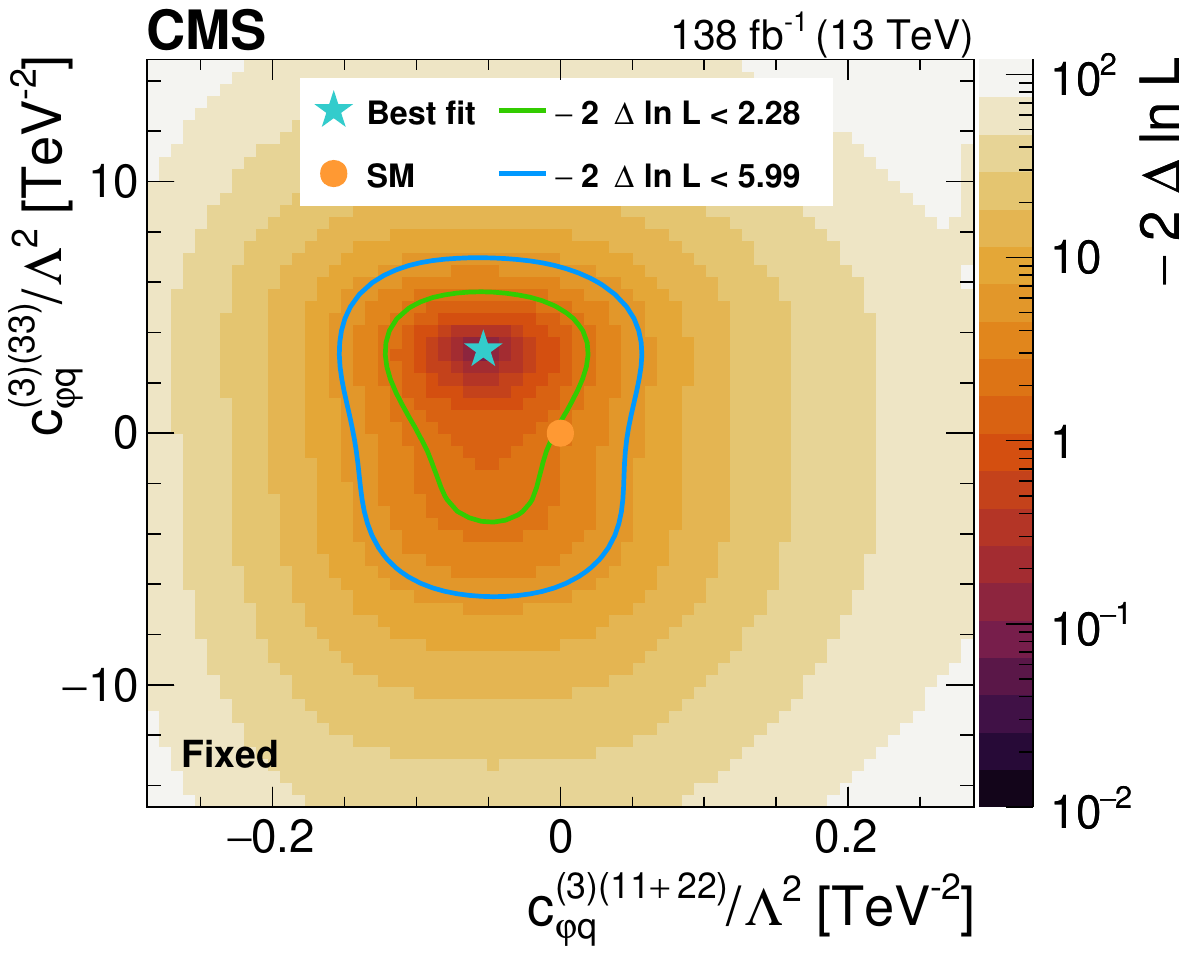}\\
    \includegraphics[width=0.49\textwidth]{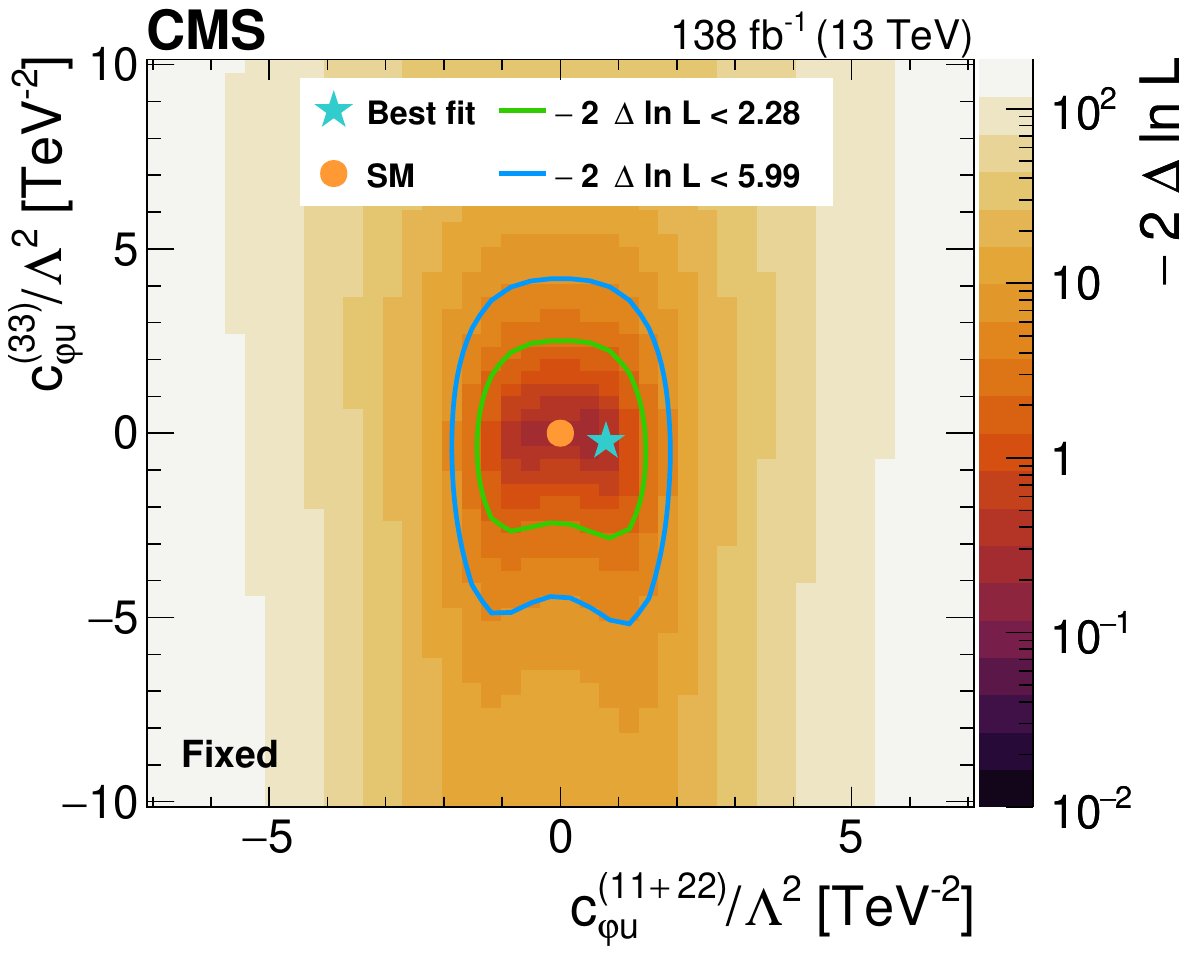}
    \includegraphics[width=0.49\textwidth]{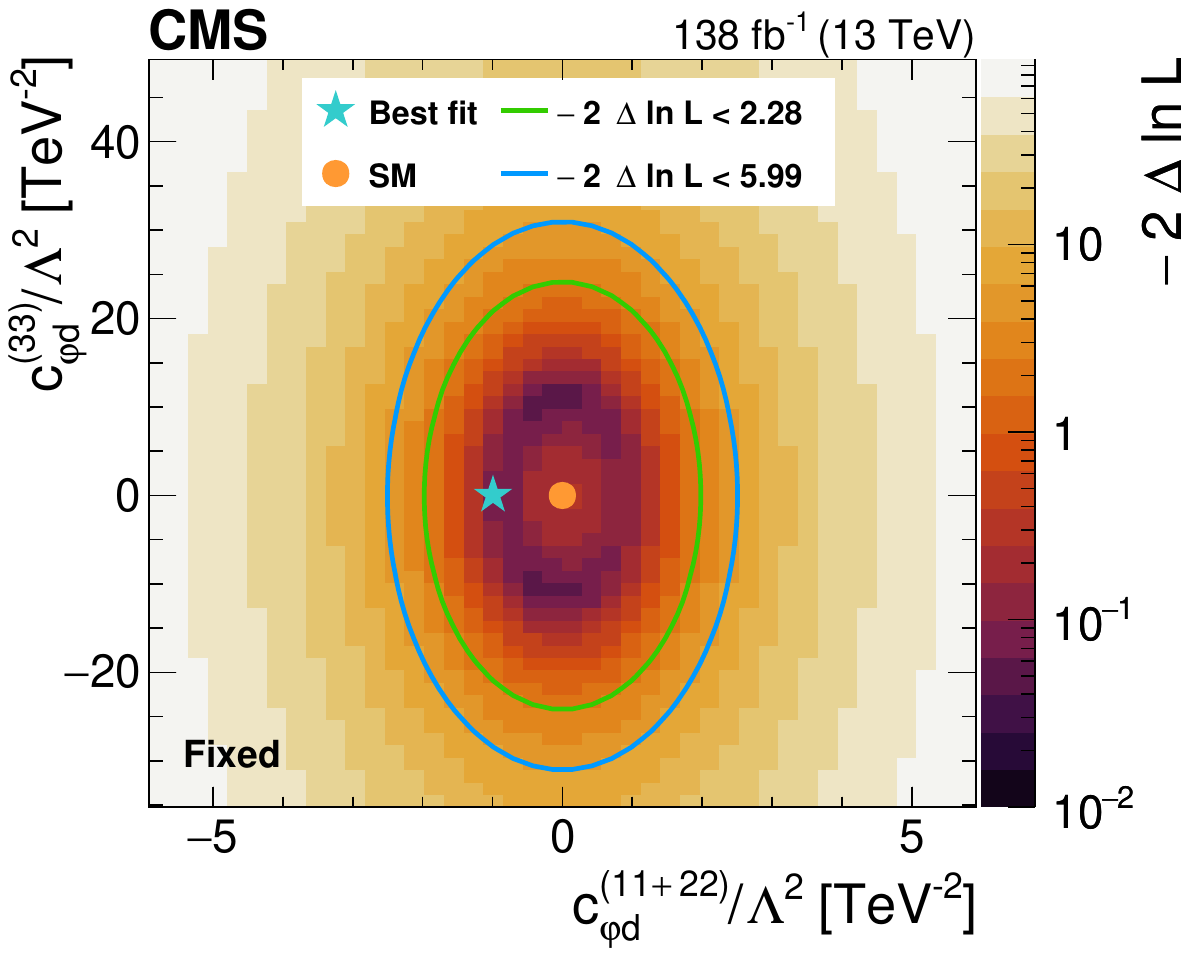} \\
    \includegraphics[width=0.49\textwidth]{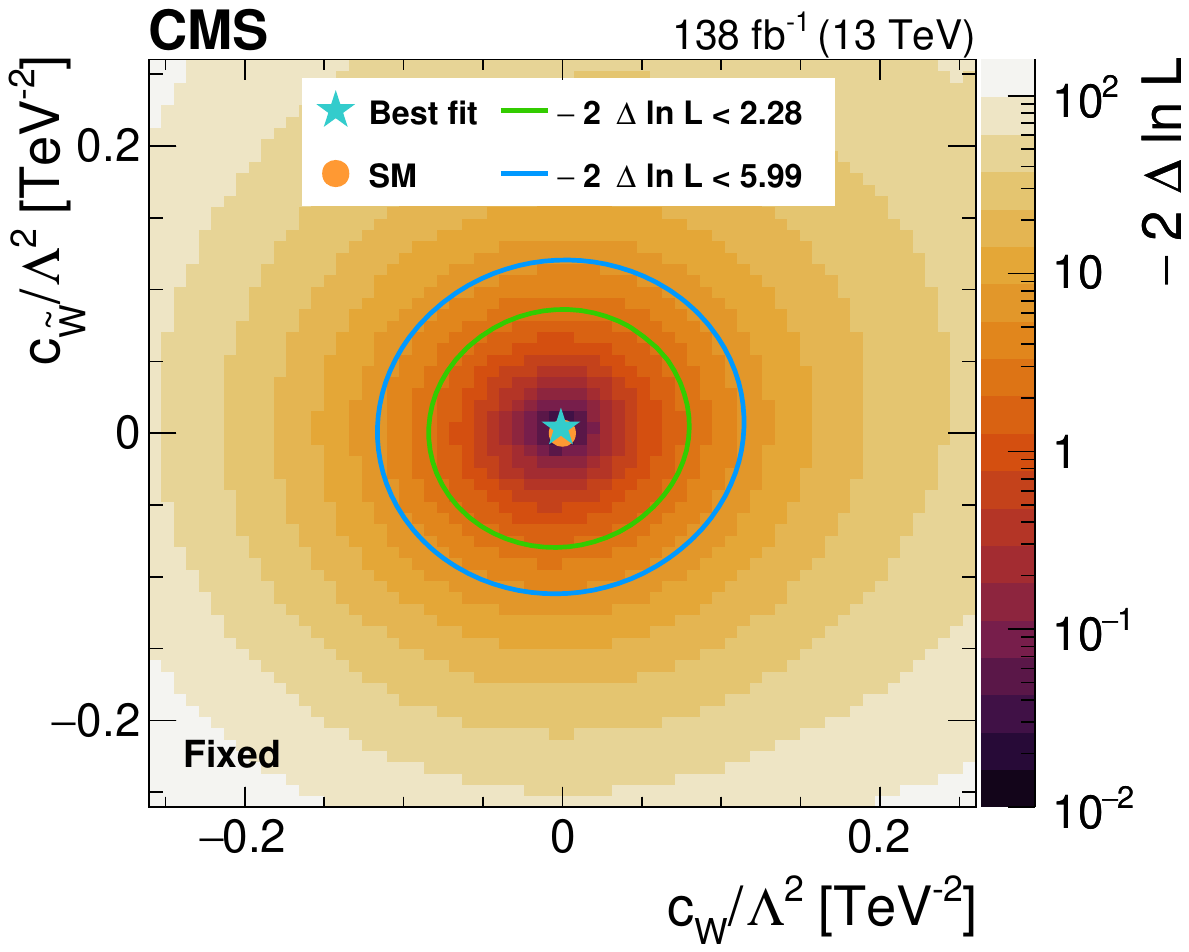}
    \caption{Likelihood as a function of the Wilson coefficients \cHqMinusLight
    and \cHqMinusThird~(upper \cmsLeft), \cHqTripletLight and \cHqTripletThird~(upper \cmsRight),
    \cHuLight and \cHuThird~(middle \cmsLeft), \cHdLight and \cHdThird~(middle \cmsRight), as well as
    \cW and \cWtil~(lower).
    All Wilson coefficients that are not scanned are fixed to zero.
    The best fit value is shown with a marker and the coloured lines correspond to the crossing points of $-2\Delta\ln L$
    at 2.28 and 5.99, which correspond to the 68 and 95\% \CL limits in the asymptotic approximation.
    \label{fig:2Dlimits}}
\end{figure*}

\begin{figure*}[tbp]
  \centering
  \includegraphics[width=0.49\textwidth]{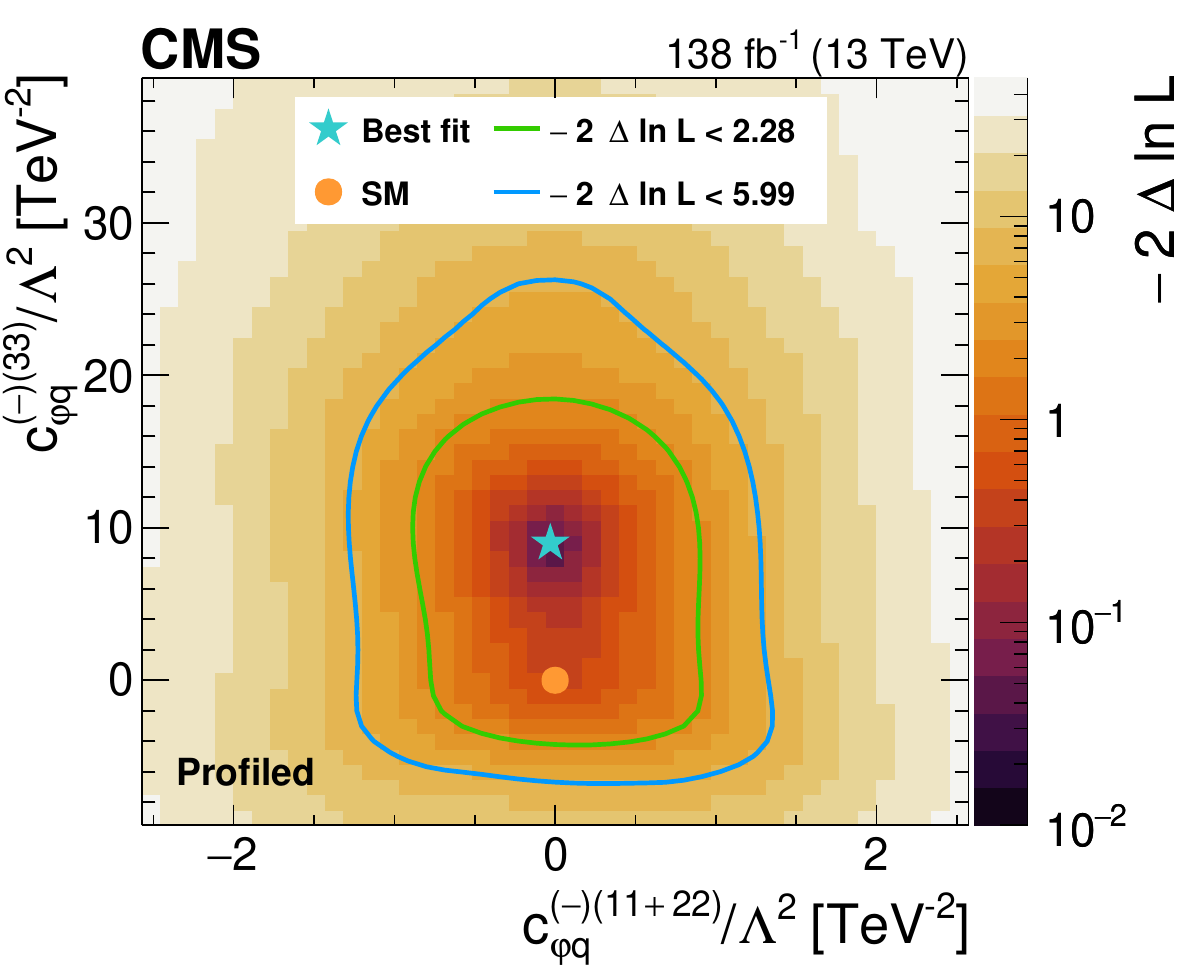}
  \includegraphics[width=0.49\textwidth]{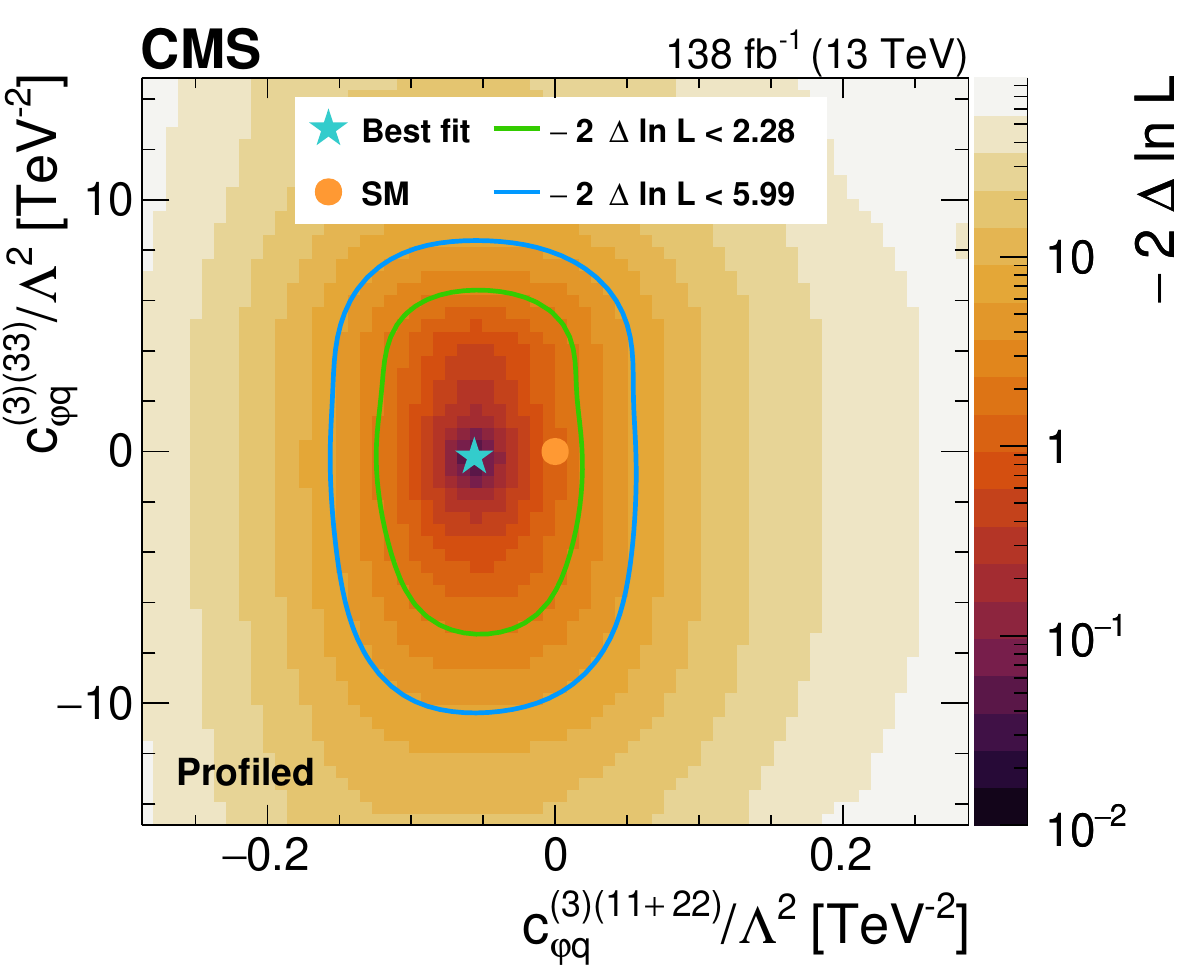}\\
  \includegraphics[width=0.49\textwidth]{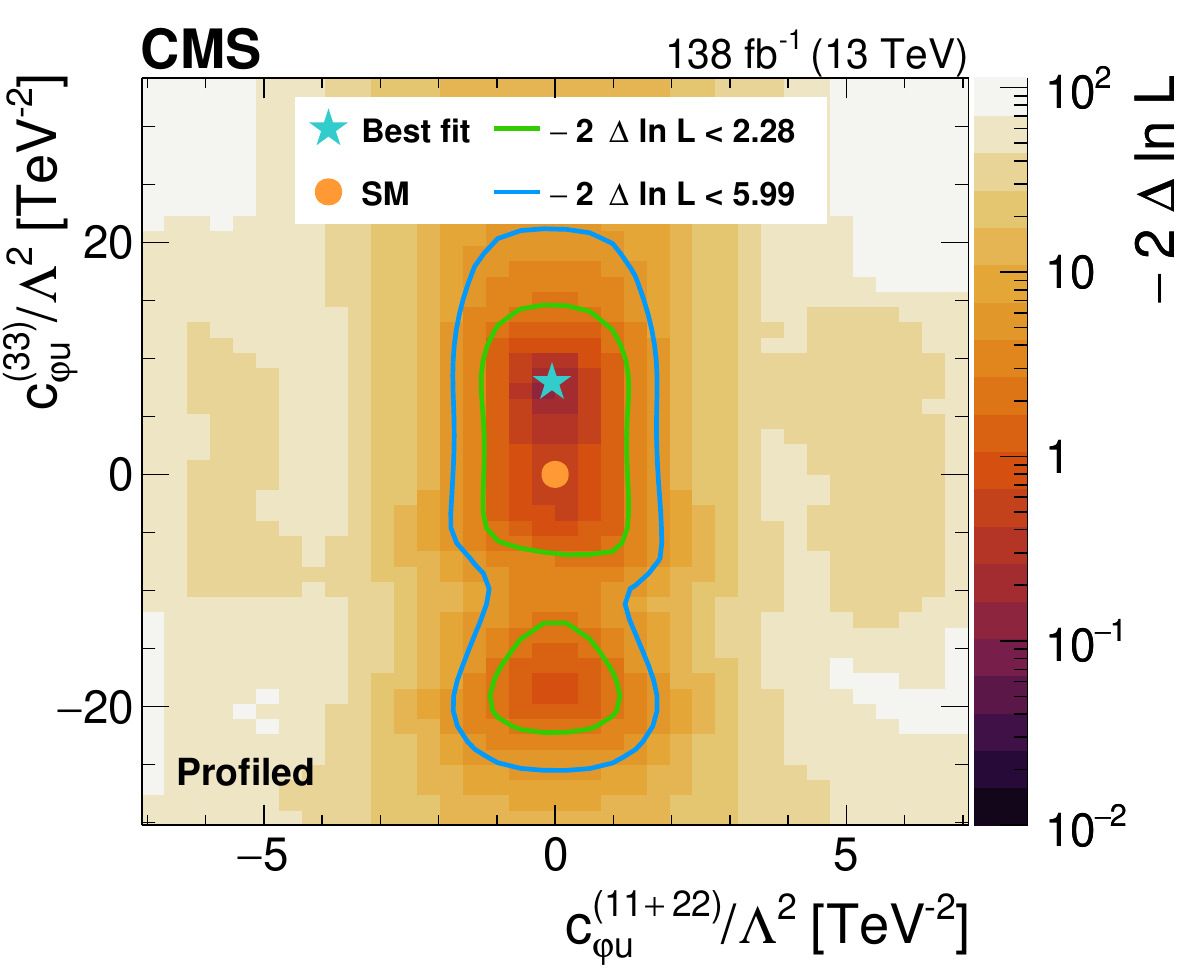}
  \includegraphics[width=0.49\textwidth]{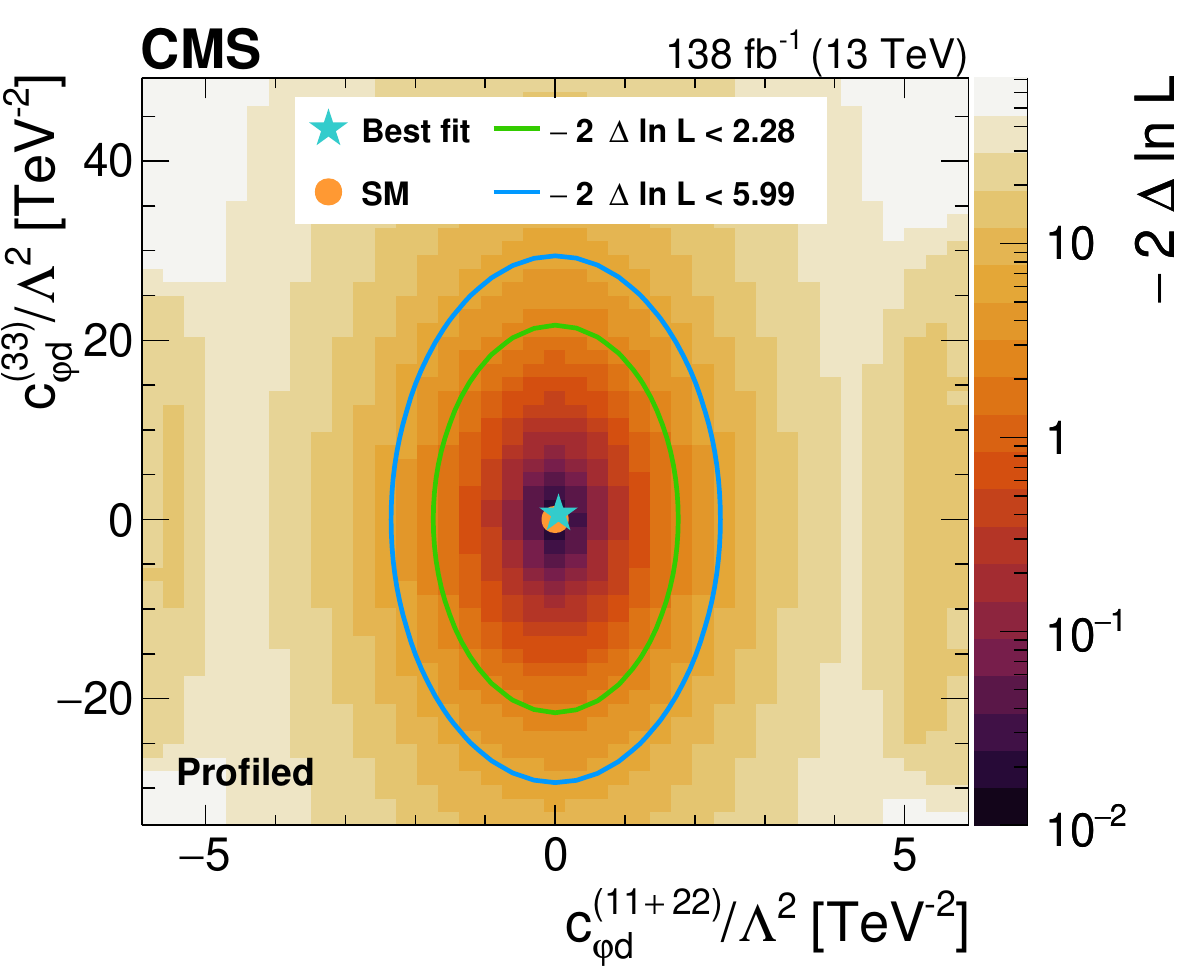} \\
  \includegraphics[width=0.49\textwidth]{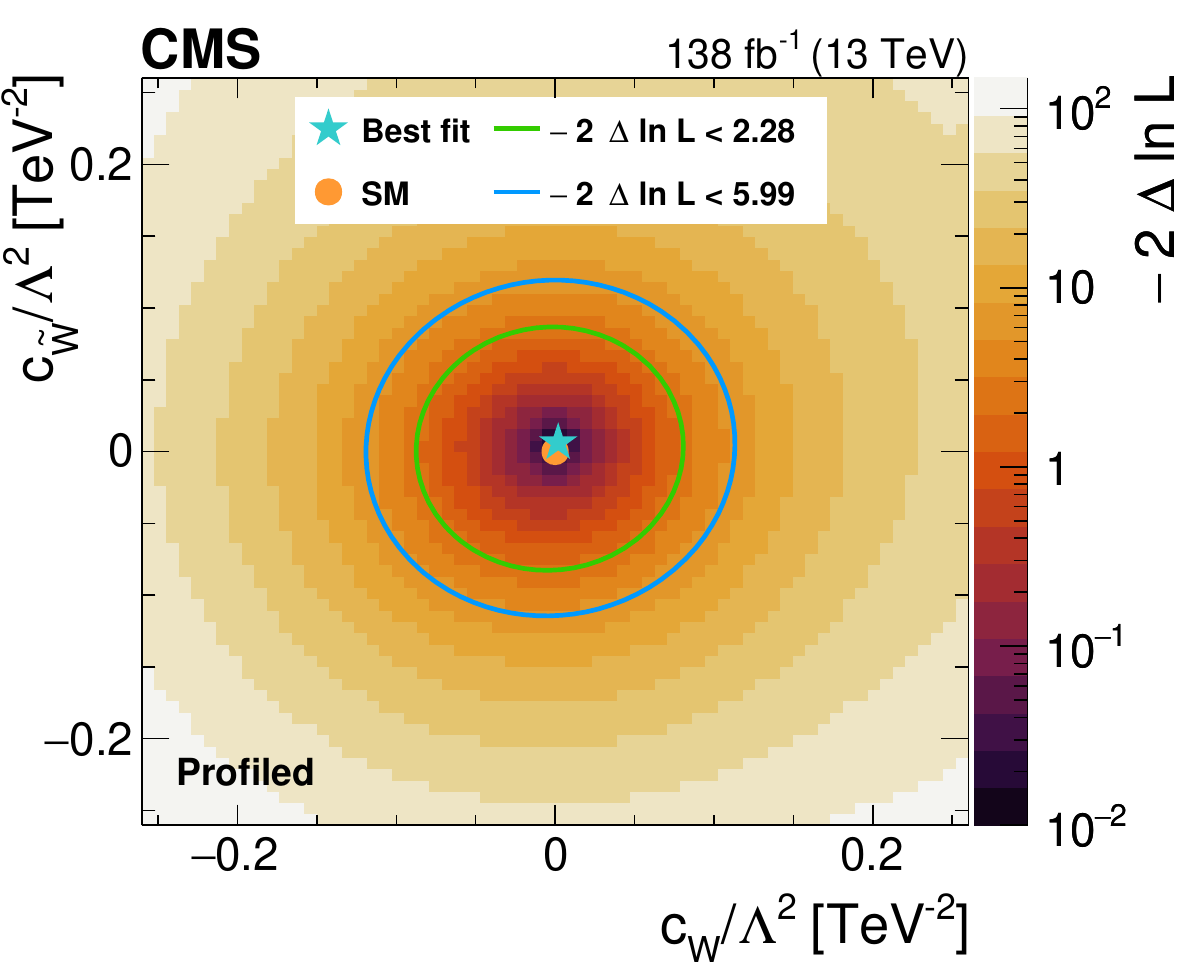}
  \caption{Likelihood as a function of the Wilson coefficients \cHqMinusLight
  and \cHqMinusThird~(upper \cmsLeft), \cHqTripletLight and \cHqTripletThird~(upper \cmsRight),
  \cHuLight and \cHuThird~(middle \cmsLeft), \cHdLight and \cHdThird~(middle \cmsRight), as well as
  \cW and \cWtil~(lower).
  Other Wilson coefficients are allowed to float in the fit.
  The best fit value is shown with a marker and the coloured lines correspond to the crossing points of $-2\Delta\ln L$
  at 2.28 and 5.99, which correspond to the 68 and 95\% \CL limits in the asymptotic approximation.
  \label{fig:2Dlimits_float}}
\end{figure*}

\begin{figure*}[tbp]
  \centering
    \includegraphics[width=0.98\textwidth]{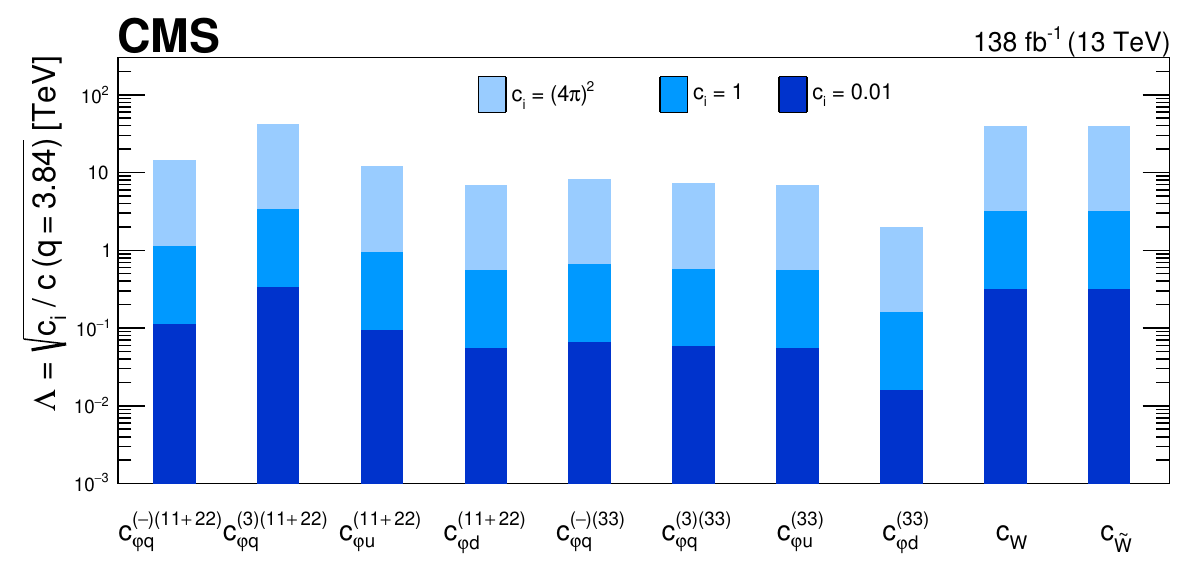}
    \caption{Summary of the limits in the energy scale $\Lambda$ obtained from the limits on the Wilson coefficients
    \cHqMinusLight, \cHqTripletLight, \cHuLight, \cHdLight, \cHqMinusThird, \cHqTripletThird, \cHuThird, \cHdThird, \cW, and \cWtil.
    Shown are limits for scans where other Wilson coefficients are fixed to zero.
    The limit on $\Lambda$ is calculated from the Wilson coefficient value at $q = -2\Delta\ln L = 3.84$,
    which corresponds to the 95\% \CL limit in the asymptotic approximation.
    The least stringent limit is chosen for each Wilson coefficient and limits are shown for three scenarios of $c_{i}$.
    \label{fig:summary_lambda}}
\end{figure*}

In order to find the best fit value of the WCs, we calculate the log-likelihood ratio
of the EFT prediction and the SM hypothesis as a function of the WCs and evaluate the difference
\begin{equation}
-2\Delta\ln L(\vec{c},\vec{\nu}) = -2(\ln L(\vec{c},\vec{\nu}) - \ln L_\text{MLE}) = q,
\end{equation}
where $\vec{c}$ are the WCs, $\vec{\nu}$ are the nuisance parameters,
$L$ is the profiled likelihood ratio and $L_\text{MLE}$ is its maximum value.
The likelihood is built from the distributions after fitting the nuisance parameters
and contains the contributions from all SRs.
Using a goodness-of-fit test based on the saturated-model likelihood~\cite{saturated} with 1000 toy experiments, 
as implemented in \textsc{Combine}, we obtain a $p$-value of 85\% for the fit shown in Fig.~\ref{fig:signalRegionsPostFit}.
The difference of the likelihood under the SM hypothesis, $L(\vec{c}=0,\vec{\nu})$, and the best fit is 
$-2(\ln L(\vec{c}=0,\vec{\nu}) - \ln L_\text{MLE}) = 2.5$.
Figure~\ref{fig:summary} shows the best fit values and the confidence intervals for all WCs.
It contains results where other WCs are fixed to zero (`fixed') as well as those with other WCs floating in the fit (`profiled').
Figures~\ref{fig:2Dlimits} and~\ref{fig:2Dlimits_float} show the two-dimensional
likelihood scans as functions of the light- and heavy-quark flavour couplings
for the fixed and profiled scenarios.
Figure~\ref{fig:summary_lambda} shows the limits of the new physics energy scale $\Lambda$ for each WC.
These limits are obtained via $\Lambda = \sqrt{c_{i}/c(q=3.84)}$,
where $c_{i}$ is a fixed value and $c(q=3.84)$ corresponds to the value of the WC at $q = -2\Delta\ln L = 3.84$.
For each WC, the least stringent limit on $\Lambda$ is chosen for display.
In the asymptotic approximation, the intervals from the crossing points of $q$ at 1 and 3.84 correspond
to the 68 and 95\% confidence-level (\CL) limits.
In the quadratic EFT model, however, Wilks' theorem is not guaranteed to hold for cases where sizeable quadratic contributions are present~\cite{Bernlochner:2022oiw}.
Only the operators \cW and \cWtil show sizable quadratic contributions, such that for all flavour dependent WCs, 
fits including only linear terms yield results consistent with those from the full EFT model and the violation of Wilks' theorem has little impact.

Overall, we observe consistency with the SM within the uncertainties.
This is further verified by extracting the production cross sections of the \ttZ, \WZ, and \ZZ
processes with all WCs set to zero.
All are compatible with the SM predictions and dedicated measurements~\cite{CMS:2020gtj,CMS:2021icx} within one standard deviation.
Compared to the SM prediction, the \WZ and \ZZ cross sections are measured 5\% lower and 10\% higher, respectively.
This effect is compatible with the slight deviations from zero in the fits of
\cHqTripletLight and \cHqTripletThird as presented in Figs.~\ref{fig:2Dlimits} and~\ref{fig:2Dlimits_float}.

The WCs \cHqMinusLight, \cHqMinusThird, \cHuLight, \cHuThird,
\cHdLight, and \cHdThird are constrained by the combination
of processes that are sensitive to the light- and heavy-quark generations.
The effects of \cHuThird and \cHqMinusThird are largely correlated and cannot
be fully disentangled in a fit with all EFT parameters.
Thus, both parameters show different sizes of uncertainties comparing the fixed and profiled versions of the fit in Fig.~\ref{fig:summary}.
In addition, cancellations between \cHuThird and \cHqMinusThird and the interplay between the linear and quadratic EFT terms
cause a secondary minimum in the two-dimensional likelihood
scan of \cHuLight and \cHuThird along the \cHuThird axis in Fig.~\ref{fig:2Dlimits_float}~(middle left).
Some WCs, especially \cW and \cWtil, are measured close to zero,
which reflects the slight deficit in the \WZ contribution and the fact that parameters with large quadratic terms can only increase the event yield.

Variations of \cHqTripletLight change the coupling of the \PW boson to quarks and
can be thoroughly probed in \WZ production.
The coupling described by \cHqTripletThird cannot be precisely probed with the
setup in this analysis.
There is some sensitivity from the \PZ-\PQb vertex in \ZZ production
but to probe this coupling more precisely, a vertex with a heavy quark and
\PW boson is needed.
Furthermore, this vertex must be a production vertex.
Even though such a vertex exists in the top quark decays of the \ttZ production,
it does not add sensitivity to \cHqTripletThird because it only alters the top quark decay
and not the production cross section.
It was verified that a simple angular analysis of the top quark decay products
does not add sensitivity.
Compared to another analysis focusing on associated \ttbar production~\cite{CMS:2023xyc},
similar sensitivity is obtained to \cHqMinusThird, when fixing other WCs to zero.
Compared to an analysis focusing on light quarks~\cite{CMS:2024ksn},
weaker limits in \cHqMinusLight but comparable limits in \cHqTripletLight are observed.
Global fits including EW precision data can further constrain \cHqMinusLight and \cHqTripletLight but obtain 
similar limits for \cW~\cite{Ellis:2018gqa, Ellis:2020unq}.
However, the presented analysis simultaneously constrains possible flavour differences,
which are usually not considered in EFT measurements.
The coefficients \cW and \cWtil can be constrained by the measurement of the \WZ process,
which can feature a $\PW\PW\PZ$ vertex.
Dedicated analyses reach higher sensitivity to these couplings~\cite{ATLAS:2021jgw,CMS:2021icx}.

\section{Summary}
\label{sec:summary}
An analysis of the flavour structures in effective field theory~(EFT) couplings has been presented, considering the electroweak coupling to quarks
of different generations in the processes \ttZ, \WZ, and \ZZ.
Proton-proton collision data, collected at $\sqrt{s}=13\TeV$
in 2016--2018 by the CMS detector and corresponding to an integrated luminosity of 138\fbinv were analysed.
For the first time, the flavour structures of the \PZ-quark couplings are disentangled by
simultaneously probing the light- and heavy-quark couplings in different processes.
The measured Wilson coefficients are compatible with the standard model hypothesis
within their uncertainties and corresponding limits are placed using one- and two-dimensional scans
of the profiled likelihood test statistic.
Extracting the EFT parameters from multiple processes simultaneously makes it possible to
correctly correlate EFT effects of the three processes that are often important backgrounds of each other.
Thus, these results contribute to a more comprehensive EFT interpretation that preserves correlations in combinations or global fits,
rather than focusing solely on individual processes.

\begin{acknowledgments}
We congratulate our colleagues in the CERN accelerator departments for the excellent performance of the LHC and thank the technical and administrative staffs at CERN and at other CMS institutes for their contributions to the success of the CMS effort. In addition, we gratefully acknowledge the computing centres and personnel of the Worldwide LHC Computing Grid and other centres for delivering so effectively the computing infrastructure essential to our analyses. Finally, we acknowledge the enduring support for the construction and operation of the LHC, the CMS detector, and the supporting computing infrastructure provided by the following funding agencies: SC (Armenia), BMBWF and FWF (Austria); FNRS and FWO (Belgium); CNPq, CAPES, FAPERJ, FAPERGS, and FAPESP (Brazil); MES and BNSF (Bulgaria); CERN; CAS, MoST, and NSFC (China); MINCIENCIAS (Colombia); MSES and CSF (Croatia); RIF (Cyprus); SENESCYT (Ecuador); ERC PRG, RVTT3 and MoER TK202 (Estonia); Academy of Finland, MEC, and HIP (Finland); CEA and CNRS/IN2P3 (France); SRNSF (Georgia); BMBF, DFG, and HGF (Germany); GSRI (Greece); NKFIH (Hungary); DAE and DST (India); IPM (Iran); SFI (Ireland); INFN (Italy); MSIT and NRF (Republic of Korea); MES (Latvia); LMTLT (Lithuania); MOE and UM (Malaysia); BUAP, CINVESTAV, CONACYT, LNS, SEP, and UASLP-FAI (Mexico); MOS (Montenegro); MBIE (New Zealand); PAEC (Pakistan); MES and NSC (Poland); FCT (Portugal); MESTD (Serbia); MICIU/AEI and PCTI (Spain); MOSTR (Sri Lanka); Swiss Funding Agencies (Switzerland); MST (Taipei); MHESI and NSTDA (Thailand); TUBITAK and TENMAK (T\"{u}rkiye); NASU (Ukraine); STFC (United Kingdom); DOE and NSF (USA).

\hyphenation{Rachada-pisek} Individuals have received support from the Marie-Curie programme and the European Research Council and Horizon 2020 Grant, contract Nos.\ 675440, 724704, 752730, 758316, 765710, 824093, 101115353, 101002207, 101001205, and COST Action CA16108 (European Union); the Leventis Foundation; the Alfred P.\ Sloan Foundation; the Alexander von Humboldt Foundation; the Science Committee, project no. 22rl-037 (Armenia); the Fonds pour la Formation \`a la Recherche dans l'Industrie et dans l'Agriculture (FRIA-Belgium); the Beijing Municipal Science \& Technology Commission, No. Z191100007219010, the Fundamental Research Funds for the Central Universities, the Ministry of Science and Technology of China under Grant No. 2023YFA1605804, and the Natural Science Foundation of China under Grant No. 12061141002 (China); the Ministry of Education, Youth and Sports (MEYS) of the Czech Republic; the Shota Rustaveli National Science Foundation, grant FR-22-985 (Georgia); the Deutsche Forschungsgemeinschaft (DFG), among others, under Germany's Excellence Strategy -- EXC 2121 ``Quantum Universe" -- 390833306, and under project number 400140256 - GRK2497; the Hellenic Foundation for Research and Innovation (HFRI), Project Number 2288 (Greece); the Hungarian Academy of Sciences, the New National Excellence Program - \'UNKP, the NKFIH research grants K 131991, K 133046, K 138136, K 143460, K 143477, K 146913, K 146914, K 147048, 2020-2.2.1-ED-2021-00181, TKP2021-NKTA-64, and 2021-4.1.2-NEMZ\_KI-2024-00036 (Hungary); the Council of Science and Industrial Research, India; ICSC -- National Research Centre for High Performance Computing, Big Data and Quantum Computing, FAIR -- Future Artificial Intelligence Research, and CUP I53D23001070006 (Mission 4 Component 1), funded by the NextGenerationEU program (Italy); the Latvian Council of Science; the Ministry of Education and Science, project no. 2022/WK/14, and the National Science Center, contracts Opus 2021/41/B/ST2/01369 and 2021/43/B/ST2/01552 (Poland); the Funda\c{c}\~ao para a Ci\^encia e a Tecnologia, grant CEECIND/01334/2018 (Portugal); the National Priorities Research Program by Qatar National Research Fund; MICIU/AEI/10.13039/501100011033, ERDF/EU, "European Union NextGenerationEU/PRTR", and Programa Severo Ochoa del Principado de Asturias (Spain); the Chulalongkorn Academic into Its 2nd Century Project Advancement Project, and the National Science, Research and Innovation Fund via the Program Management Unit for Human Resources \& Institutional Development, Research and Innovation, grant B39G680009 (Thailand); the Kavli Foundation; the Nvidia Corporation; the SuperMicro Corporation; the Welch Foundation, contract C-1845; and the Weston Havens Foundation (USA).
\end{acknowledgments}

\bibliography{auto_generated}
\cleardoublepage \appendix\section{The CMS Collaboration \label{app:collab}}\begin{sloppypar}\hyphenpenalty=5000\widowpenalty=500\clubpenalty=5000\cmsinstitute{Yerevan Physics Institute, Yerevan, Armenia}
{\tolerance=6000
A.~Hayrapetyan, V.~Makarenko\cmsorcid{0000-0002-8406-8605}, A.~Tumasyan\cmsAuthorMark{1}\cmsorcid{0009-0000-0684-6742}
\par}
\cmsinstitute{Institut f\"{u}r Hochenergiephysik, Vienna, Austria}
{\tolerance=6000
W.~Adam\cmsorcid{0000-0001-9099-4341}, J.W.~Andrejkovic, L.~Benato\cmsorcid{0000-0001-5135-7489}, T.~Bergauer\cmsorcid{0000-0002-5786-0293}, M.~Dragicevic\cmsorcid{0000-0003-1967-6783}, C.~Giordano, P.S.~Hussain\cmsorcid{0000-0002-4825-5278}, M.~Jeitler\cmsAuthorMark{2}\cmsorcid{0000-0002-5141-9560}, N.~Krammer\cmsorcid{0000-0002-0548-0985}, A.~Li\cmsorcid{0000-0002-4547-116X}, D.~Liko\cmsorcid{0000-0002-3380-473X}, M.~Matthewman, I.~Mikulec\cmsorcid{0000-0003-0385-2746}, J.~Schieck\cmsAuthorMark{2}\cmsorcid{0000-0002-1058-8093}, R.~Sch\"{o}fbeck\cmsAuthorMark{2}\cmsorcid{0000-0002-2332-8784}, D.~Schwarz\cmsorcid{0000-0002-3821-7331}, M.~Shooshtari, M.~Sonawane\cmsorcid{0000-0003-0510-7010}, W.~Waltenberger\cmsorcid{0000-0002-6215-7228}, C.-E.~Wulz\cmsAuthorMark{2}\cmsorcid{0000-0001-9226-5812}
% ...
\par}
\vskip\cmsinstskip
\dag:~Deceased\\
$^{1}$Also at Yerevan State University, Yerevan, Armenia\\
$^{2}$Also at TU Wien, Vienna, Austria\\
% ...
\end{sloppypar}
\end{document}